\documentclass[11pt]{article}
\pdfoutput=1
\usepackage{bbold}
\usepackage{amsmath}
\usepackage{url}
\usepackage{hyperref}
\usepackage{slashed}
\usepackage{tikz}
\usetikzlibrary{decorations.pathmorphing}
\usetikzlibrary{matrix}
\usepackage{graphicx}
\usepackage{epstopdf}
\usepackage{subfigure}
\usepackage{pgfplots}
\usepackage{amssymb}
\usepackage{color}
\usepackage{mathrsfs}
\usepackage{fancybox}
\usepackage{lipsum}
\usepackage{eurosym}
\usepackage{tcolorbox}
\usepackage{anyfontsize}
\usepackage{enumitem,kantlipsum}

\newcommand{\Tr}{\mbox{Tr}}

\usetikzlibrary{decorations.pathmorphing}
\tikzset{snake it/.style={decorate, decoration=snake}}
\let\originallesssim\lesssim
\let\originalgtrsim\gtrsim
\DeclareRobustCommand{\lesssim}{%
  \mathrel{\mathpalette\lowersim\originallesssim}%
}
\DeclareRobustCommand{\gtrsim}{%
  \mathrel{\mathpalette\lowersim\originalgtrsim}%
}

\makeatletter
\newcommand{\lowersim}[2]{%
  \sbox\z@{$#1<$}%
  \raisebox{-1pt}{\small $\m@th#1#2$}%
}
\def\({\left (}
\def\){\right )}
\def\[{\left [}
\def\]{\right ]}

\numberwithin{equation}{section}

\usepackage{environ}
\NewEnviron{myequation}{%
\begin{eqnarray}
\scalebox{1.05}{$\BODY$}
\end{eqnarray}
}   
\newcommand{\beq}{\begin{equation}}
\newcommand{\eeq}{\end{equation}}
 
\newcommand{\bea}{\begin{eqnarray}}
\newcommand{\ea}{\end{eqnarray}}
\newcommand{\barr}{\!\begin{array}}
\newcommand{\earr}{\end{array}\!}

\usepackage{mciteplus} 
\def\nspc{\hspace{-.5pt}}

\def\ddelta{{{\mbox{\fontsize{7pt}{7pt}${\nspc\Delta}$}}}}
\def\spc{\hspace{1pt}}

\counterwithout{equation}{section}

\def\smpc{{\hspace{.5pt}}}
\usepackage{upgreek}
\def\lambdabar{{\frac \lambda {2\pi}}}
\def\lambdabar{\lambda}

\usetikzlibrary{decorations.pathreplacing,decorations.markings,snakes}

\def\be{\begin{equation}}
\def\ee{\end{equation}}

\def\la{\langle}
\def\bea{\begin{eqnarray}}
\def\eea{\end{eqnarray}}
\def\is{\!  & \!  = \!  &  \!}
\def\ra{\rangle}

\def\ea{\eea}

\def\qbigrr{,\smpc q\bigr)}

\def\be{\bea}
\def\ee{\eea}

\renewcommand\large{\fontsize{13}{13.5}\selectfont}
\def\Aminus{\,\overline{\!\cal A}\spc}
\def\Aplus{{{\cal A}\spc}}

\def\spc{\hspace{1pt}}
\def\is{\! &  \! = \! & \!}

\def\ttau{\mbox{\scriptsize${T}\nspc$}}
\def\rmC{{\rm C}}
\def\mfab{\alpha}

\def\MM{{\cal M}}
\def\lL{{{\mbox{\fontsize{7pt}{7.5pt}${L}$}}}}
\def\rR{{{\mbox{\fontsize{7pt}{7.5pt}${R}$}}}}
\usepackage{titling} 
\usepackage[affil-it]{authblk} 
\begin{document}

\addtolength{\topmargin}{3cm}

\title{\bf SYK Correlators from 2D Liouville-de Sitter Gravity}
\author[1,2]{Herman Verlinde}
\author[1]{Mengyang Zhang}%

\affil[1]{ Department of Physics, Princeton University, Princeton, NJ 08544}

\medskip

\affil[2]{School of Natural Sciences, Institute for Advanced Study, Princeton, NJ 08540}

    \maketitle
\bigskip

\bigskip

    \begin{abstract}
   {We introduce and study a candidate gravity dual to the double scaled SYK model in the form of an exactly soluble 2D de Sitter gravity model consisting of two spacelike Liouville CFTs with complex central charge adding up to $c_+ + c_- = 26$. In \cite{ustwo} it was shown that the two-point function of physical operators in a doubled SYK model matches in the semi-classical limit with the Green's function of a massive scalar field in 3D de Sitter space. As further evidence of the duality, we adapt a result from Zamolodchikov to compute the boundary two-point function of the 2D Liouville-de Sitter gravity model on a disk and find that it reproduces the exact DSSYK two-point function to all orders in $\lambda=p^2/N$. We describe how the 2D Liouville-de Sitter gravity model arises from quantizing 3D de Sitter gravity.}
    \end{abstract}

\addtolength{\topmargin}{-3cm}
\pagebreak

\tableofcontents
\addtolength{\baselineskip}{0.5mm}
\addtolength{\parskip}{1mm}
\addtolength{\abovedisplayskip}{.65mm}
\addtolength{\belowdisplayskip}{.65mm}
\pagebreak

\def\qbigrt{\bigr)}
\def\darkblue{blue!90!black}
\def\darkredd{red!90!black}
\def\darkred{black}
\def\darkgreen{green!50!black}
\def\lcyan{cyan!50!white}

\def\lgray{gray!50!white}
\section{Introduction}

\vspace{-1mm} 

The SYK model has been actively studied recently as an example of a quantum chaotic model with dynamical properties that motivate a correspondence with low dimensional gravity  \cite{kitaevTalks,Maldacena:2016hyu, Berkooz:2018jqr,Berkooz:2018qkz, Mukhametzhanov:2023tcg, Lin:2022rbf, Lin:2023trc}. At low temperature, the model reduces to Schwarzian quantum mechanics and furnishes a plausible holographic dual to near-AdS${}_2$ gravity \cite{Almheiri:2014cka,Jensen:2016pah,Maldacena:2016upp, Engelsoy:2016xyb, Mertens:2017mtv,Iliesiu:2019xuh}. However, there are several hints that the high temperature limit of the double scaled SYK model can provide a quantum description of low-dimensional de Sitter space \cite{HVtalks}\cite{Susskind:2021esx,Susskind:2022bia, Susskind:2022dfz, Lin:2022nss}\cite{Rahman:2022jsf}\cite{ustwo}\cite{Verlinde:2024znh}. 

In \cite{ustwo} it was shown that there is a precise match between the two-point function of physical SYK operators evaluated between two maximal entropy states and the scalar Green's function in 3D de Sitter space. The construction involved a pair of SYK models with identical Hamiltonians 
\bea
H_{L,R} =  \, i^{p/2} \! \sum_{i_1\ldots i_p} \! J^{\rR,\lL}_{\, i_1\ldots i_p} \psi^{\lL,\rR}_{i_1} \ldots \psi^{\lL,\rR}_{i_p}\,,
\eea
coupled via the equal energy constraint 
on physical states and operators
\bea
\label{zerotwo}
(H_L \!\spc - H_R)|{\rm \smpc phys \smpc} \ra\!  \is\! 0 \qquad \qquad \bigl[H_L\! -H_R, {\cal O}_\Delta^{\rm phys}\bigr]\spc =\spc 0
\eea
This coupled model is soluble in the double scaled limit $N,p \to \infty$ with $\lambda = 2p^2/N$ fixed. It has the same energy spectrum as a single double scaled SYK model. In particular, there is a special energy eigenstate $| E_0\ra = | E_0\ra_L | E_0\ra_R$ 
with  maximal entropy and infinite temperature, in the sense that it sits at the point in the spectrum where the spectral density $\rho(E) = e^{S(E)}$ has a maximum as a function of $E$. We will denote this state by
\bea
|\Psi_{\rm dS} \ra \is | E_0\ra, \qquad \qquad E_0 = 0
\eea 
In \cite{ustwo} it was proposed that this maximal entropy state $|\Psi_{\rm dS} \ra$ describes the vacuum state in a 2D gravity model given by the circle reduction of 3D de Sitter gravity. A similar proposal was made in \cite{Susskind:2021esx}\cite{Rahman:2023pgt} based on other pieces of evidence and with a different identification of parameters.

 Starting from the exact expression for the two point function of two such physical operators ${\cal O}^+_\Delta(\tau)$ and ${\cal O}^-_\Delta(0)$ separated by a time-distance $\tau$  one finds that in the semi-classical large $N$ limit it takes the form \cite{ustwo}
\bea
\label{gab}
G^{+-}_{\! \Delta}(\tau) \!\is\! \bigl\la \Psi_{\rm dS} \bigl| \spc {\cal O}^+_{\!\Delta} (\tau) \spc {\cal O}^-_{\!\Delta} (0)\spc  \bigr|\Psi_{\rm dS} \bigr\ra{\strut}{\strut} \
\spc = \spc \frac{2\sinh\bigl( \mu(i\pi + \tau)\bigr)\strut}{\pi \sinh \tau {\strut}}\
\eea
with $\mu = 2\Delta-1$. The right-hand side coincides with the Green's function of a scalar field of mass $m^2 = 4\Delta(1-\Delta)$ evaluated at two points $x_1$ and $x_2$ separated by a geodesic distance $\tau = \tau(x_1,x_2)$ in 3D de Sitter space \cite{Bousso:2001mw}. This correspondence points to an interpretation of the high temperature limit of the double scaled SYK model as a microscopic dual to low dimensional de Sitter gravity.

It is natural to hope that this match is just one element of a larger holographic dictionary. The double scaled SYK indeed has several other properties that point to a link with 2D and 3D de Sitter gravity. First, it has a bounded energy spectrum labeled by an angle $\theta$. This mirrors the fact that a localized mass $M$ in 3D de Sitter space produces a conical singularity with an angle deficit $\theta$ proportional to $M$. 
SYK correlation functions furthermore exhibit a $U_{\mathsf q}(\mathfrak{sl}_2)$ quantum group symmetry \cite{Berkooz:2018jqr,Lin:2023trc} with $\mathsf{q} \in [0,1]$ that looks identical to the the quantum group that governs the braiding properties of point masses  in 3D de Sitter gravity or conformal blocks in 2D de Sitter gravity \cite{Verlinde:2024znh}. Finally, the collective field theory of DSSYK looks identical to the edge CFT associated with 3D de Sitter gravity \cite{Klemm:2002ir}\cite{Cotler:2019nbi} as derived from its reformulation in terms of $SL(2,\mathbb{C})$ Chern-Simons theory \cite{Witten:1988hc,Witten:1989ip}. In this paper we will exploit these formal link to extend  the match between the two-point functions found in \cite{ustwo} to an exact equality for all finite value of the Newton constant. This result provides further evidence in support of an exact dual equivalence between these models.

This paper is organized as follows. In section 2 we review the doubled SYK model and the exact expression of its correlation functions. In section 3 we summarize the 3D de Sitter gravity and its formulation in terms of $SL(2,\mathbb{C})$ Chern-Simons theory. In sections 4 we introduce the 2D Liouville-de Sitter gravity model defined by the sum of two space-like Liouville actions with total charge $c_+ + c_- = 26$. Correlation functions of this 2D gravity theory describes the Hilbert states of 3D de Sitter gravity. We establish a precise quantitative correspondence between the doubled SYK model and the 2D Liouville-de Sitter gravity by matching the DSSYK two-point function obtained in \cite{ustwo} with the boundary two-point function of the Liouville model, building on an old result  by Zamolodchikov \cite{Zamolodchikov:2005fy}. We end with some concluding comments. 

\def\iseqto{ \!&\!\equiv\!&\! }

\section{Some Properties of Double Scaled SYK}
\vspace{-1.5mm}

We start with a short overview of some relevant properties of the double scaled SYK model and the doubled version introduced in \cite{ustwo}. As we will see in the following sections, the doubled SYK model has several properties that point to dual formulation in terms of de Sitter gravity.  

Consider a pair of SYK models \eqref{zerotwo} coupled via the equal energy constraint \eqref{zerotwo}. The condition \eqref{zerotwo} has solutions when the two SYK Hamiltonians $H_L$ and $H_R$ have equal random couplings $J^{\lL}_{\, i_1 \smpc . . . \smpc i_p} = J^{\rR}_{\, i_1 \smpc . . . \smpc i_p}.$ The solutions are linear sums of paired energy eigenstates of the total Hamiltonian 
\bea
|E \ra \! = \! \spc | E\smpc \ra_{\! L} \, | \smpc E\smpc \ra_{\! R} \qquad\   H |E\ra =   E |E\smpc \ra, \qquad \  H = \frac 1 2 ( H_{L}+ H_R)
\eea
Local SYK operators of the $L$ and $R$ model are defined via
\bea
\label{scalingo}
{\cal O}^{\lL,\rR}_\Delta \! \is \!  i^{p'/2} \sum_{i_1,\ldots,i_{p'}} K_{i_1i_2\ldots i_{p'}}\psi_{i_1}\psi_{i_1}\ldots\psi_{i_{p'}}
\eea
where $K_{i_1i_2\ldots i_{p'}}$ denotes a suitably normalized gaussian random coupling and $\Delta = p'/p$. Physical operators that preserve the equal energy constraint  \eqref{zerotwo} are given by suitable convolution integrals of a product of two local SYK operators.  A natural class of physical observables are operators of the form \cite{ustwo}
\bea
\label{opm} 
 {\cal O}^{\rm phys}_{\!\Delta}(\tau)\is \int \! dt\, {\cal O}^\lL_{1 - \Delta}(t)\spc {\cal O}^\rR_{\Delta}(-t+\tau\spc)
\eea
The integral over $t$ guarantees that ${\cal O}^{\rm phys}_{\!\Delta}(\tau)$ commutes with $H_L-H_R$.

Correlation functions of physical operators in the doubled model can be obtained by using known SYK results. SYK correlators can be represented as trivalent diagrams obtained by connecting propagators by means of three-point vertices. 
 The vertices represent the matrix elements of the local operators between energy eigenstates and the propagators represent the time evolution in the energy eigen basis. 
 They are given by the following expressions\cite{Berkooz:2018jqr,Lin:2023trc, Mukhametzhanov:2023tcg}\footnote{Here we use the short-hand notation $f(a\pm b \pm c) = f(a+b+c)f(a-b+c)f(a+b-c)f(a-b-c)$, etc.}
\bea
\label{frulestwo}
\begin{tikzpicture}[scale=0.51, baseline={([yshift=-0.22cm]current bounding box.center)}]
\draw[thick] (-0.2,0) arc (160:20:1.63);
\draw[fill,\darkblue] (-0.2,0.0375) circle (0.1);
\draw[fill,\darkblue] (2.8,0.0375) circle (0.1);
\draw (3.3, -0.2) node {\footnotesize $\tau_0$};
\draw (-0.6,-0.2) node {\footnotesize $\tau_1$};
\draw (1.25, 1.4) node {\footnotesize $s$};
\end{tikzpicture}\hspace{-2mm} \is e^{iE(s) \spc (\tau_1-\tau_0)}, \quad \quad E = -\frac{2\cos(\lambda s)}{\sqrt{\lambda(1-q)}} . \\[5mm]
\begin{tikzpicture}[scale=0.65, rotate=-90,baseline={([yshift=-0.07cm]current bounding box.center)}]
\draw[thick] (-.58,1.25) arc (50:-50:1.65);
\draw[fill,\darkblue] (0,0) circle (0.08);
 \draw[thick,\darkblue](-1.35,0) -- (0,0);
\draw (.2,-0.8) node {\footnotesize $\textcolor{black}{s_1}$};
\draw (.2,0.8) node {\footnotesize$\textcolor{black}{s_0}$};
\draw (-1,.3) node {\small\textcolor{\darkblue}{$\Delta$}};
\end{tikzpicture}\ \ \is \la E_1| {\cal O}_\Delta | E_0 \ra \ = \ \sqrt{\, \frac{\Gamma_{q}(\Delta\pm is_0\pm is)}{\Gamma_q(2\Delta)}}_{\strut}
 \eea
Here $\Gamma_q(x)$ denotes the q-Gamma function defined in Appendix A. 
The total expression obtained via these rules is integrated over the intermediate spectral parameter $s$ with the spectral measure 
 \bea
\label{sspectrum}
\rho(s) \spc = \spc\frac{1}{\Gamma_q(\pm 2is)} 
\eea
Note that the spectral measure has an approximate gaussian form with a maximum at $E_0 = 0$ and that the range of the energy spectrum is bounded \cite{Berkooz:2018qkz}. The diagrammatic rules for the doubled SYK theory are given by replacing the vertex factors by the product of the left and right vertex factors, while the propagators and spectral measure remain unchanged. 
\def\phip{\phi_{\! {}_+  }}
\def\phim{\phi_{\! {}_-  }}

Let $s_0$ denote the spectral parameter of the maximal entropy state $|\Psi_{\rm dS}\ra = |E_0\ra$ with $E_0=0$.
Applying the above rules to the two-point function of two physical operators between two maximal entropy states
of the doubled model gives
\bea
\label{ggraph}\nonumber
G_\Delta(\tau_1) \! \is  \bigl\la E_0 \bigl| \spc {\cal O}_{\!\Delta} (\tau) \spc {\cal O}_{\!\Delta} (0)\spc  \bigr|E_0\ra \\[3mm]
\is  \!\int\! ds_1\, \rho(s_1) \, e^{-\tau_1 E(s_1)} 
 \raisebox{2pt}{
 \begin{tikzpicture}[scale=0.56, rotate=-90, baseline={([yshift=-.07cm]current bounding box.center)}]
\draw[thick] (0,0) circle (1.5);
\draw[thick,\darkblue] (1.5,.05) -- (.35,.05);
\draw[thick,\darkredd] (1.5,-.05) -- (.35,-.05);
\draw[thick,\darkblue] (-1.5,.05) -- (-.35,.05);
\draw[thick,\darkredd] (-1.5,-.05) -- (-.35,-.05);
\draw[blue,fill=purple] (-1.5,.0) circle (0.12);
\draw (0,1.95) node {\footnotesize $s_1$};
\draw[blue,fill=purple] (1.5,0) circle (0.12);
\draw (0.05,-1.95) node {\footnotesize $s_0$};
\draw (-0,0) node {\footnotesize \textcolor{\darkblue}{$\Delta$}};
\end{tikzpicture}\, }  \label{gexacto} \\[-0mm]\nonumber\\[2.8mm]\nonumber
 \! \is \!\int\! ds_1\, e^{-\tau_1 E(s_1)} \;  \frac{\Gamma_{q}(\Delta\pm is_0\pm is_1)\, \Gamma_{q}(1\!-\!\Delta\pm is_0\pm is_1)}{\Gamma_q(\pm 2is) \Gamma_q(2\Delta)\Gamma_q(2\!-\!2\Delta){}_{{}^{\small \strut}}}\,
\eea

It is useful rewrite the above formula in a way that is more suitable for the comparison with 2D de Sitter gravity. Using the identity \cite{Askey} 
\bea 
\Gamma_{q \spc}(x) \Gamma_{q \spc}(1-x)\! \is \! \frac{iq^{1/8}(1-q)(q;q)^3}{q^{x/2}\vartheta_1(\frac{i\lambda}{2\pi} x,q)}
\eea
with $\lambda = -\log q$ we can re-express the right-hand side of \eqref{gexacto} in terms Jacobi theta functions. Theta functions $\vartheta_1(z,q)$ are well-known to be doubly (quasi) periodic in the argument $z$. In particular it has in infinite set of zeroes that form a regular lattice on the complex plane. The q-Gamma functions on the left-hand side each have an infinite set of poles arranged in half a regular lattice.  A pictorial representation of the above identity is given in figure 1. 

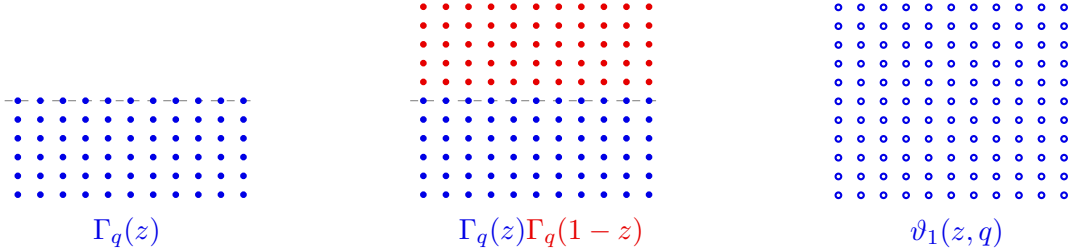
\begin{figure}
\centering
\begin{tikzpicture}[scale=.25,rotate=0]
  
 \foreach \y in {6} {
 \draw[gray, dashed] (.5,\y) -- (13.8,\y);
}
\foreach \x in {1,2,...,11} {
    \foreach \y in {1,2,...,6} {
      \draw[thick,fill=\darkblue,\darkblue] (1.2 *\x,\y) circle (0.12); 
    }
    }
    \draw (7,-1) node {\textcolor{\darkblue}{$\Gamma_q(z)$}};
    \end{tikzpicture}~~~~~~~~~~~~~~~~\begin{tikzpicture}[scale=.25,rotate=0]
  
 \foreach \y in {6} {
 \draw[gray, dashed] (.5,\y) -- (13.8,\y);
}
\foreach \x in {1,2,...,11} {
    \foreach \y in {1,2,...,6} {
      \draw[thick,fill=\darkblue,\darkblue] (1.2 * \x,\y) circle (0.12); 
    }
    }
\foreach \x in {1,2,...,11} {
    \foreach \y in {7,8,...,11} {
      \draw[thick,fill=\darkredd,\darkredd] (1.2 * \x,\y) circle (0.12); 
    }
    }
    \draw (8,-1) node {\textcolor{\darkblue}{$\Gamma_q(z)$}\textcolor{\darkredd}{$\Gamma_q(1-z)$}};
\end{tikzpicture}~~~~~~~~~~~~~~~~~~\begin{tikzpicture}[scale=.25,rotate=0]
\foreach \x in {1,2,...,11} {
    \foreach \y in {1,2,...,11} {
      \draw[thick,\darkblue] (1.2 * \x,\y) circle (0.15); 
    }
    }
    \draw (7.5,-1) node {\textcolor{\darkblue}{$\vartheta_1(z,q)$}};
\end{tikzpicture}
\vspace{-2mm}
\caption{The product of two $q$-Gamma functions has the same regular lattice of poles as the inverse power of a theta function.}
\vspace{-2mm}
\end{figure}

The SYK spectral density is expressed as a single theta function
\bea
\label{espectrum}
\rho(E) \!\is \! \rho_0 \, { \vartheta_1\bigl(\textstyle\frac{\lambda}\pi s, q\bigr) } 
\eea
and the two-point function \eqref{gexacto} of physical operators of the doubled model becomes \cite{ustwo}
\bea 
\label{gexact}
G_\Delta(\tau_1) \! \is \!\int\! dE(s_1)\, e^{-\tau_1 E(s_1)} \; \frac{\vartheta_1  \bigl( \frac{i\lambdabar}\pi \Delta \qbigrr \prod_a  \vartheta_1  \bigl( \frac{\lambdabar}\pi s_{a} \qbigrr  \strut}
{\smpc \vartheta_1  \bigl(\frac{\lambda}{2\pi} (i\Delta \pm  s_0\pm s_1)\qbigrr \strut} .
\eea
with $a\in \{0,1\}$. As we will see, the specific combination of theta functions on the right-hand side has several special properties that point to an interpretation of $G_\Delta(\tau)$ as a two-point function in 2D and 3D quantum gravity. 

The formula \eqref{gexact} played a key role in \cite{ustwo}, where it was shown that in the $q\to 1$ limit it reduces to the anti-podal scalar Green function in 3D de Sitter space. The aim of this paper is to show that this correspondence persists away from the semi-classical limit and that the above exact SYK two-point correlator matches with the gravitationally dressed two-point function in 3D de Sitter gravity with finite Newton constant $G_N=\lambda/8\pi$.  This match is a strong hint that the link between double scaled SYK and de Sitter gravity is true quantum mechanical holographic duality that extends beyond the semi-classical regime.

\section{Some Properties of 3D de Sitter Gravity}

In this section, we consider pure 3D Einstein gravity with positive cosmological constant 
on a 3D space-time manifold $\MM{}$ with boundary ${\partial M}$
\bea
\label{gravz} 
I[g] \! \! \is\!\! \frac{1}{16\pi G_3}{}_{\strut} \int^{}_{\! \MM{}\nspc}\!\! {\textstyle\sqrt{\mbox{\small$-$}{g}}}\, \bigl({R} - \Lambda \bigr) + \frac{1}{8\pi G_3} \int_{\partial M}\!\!\!  {\textstyle{\sqrt{\mbox{\small$-$}h}}}\, K 
\eea
We will give a brief overview of its first order formulation and of the identification between the 3D de Sitter gravity wave functions and the chiral partition functions of 2D conformal field theory. Our eventual goal is to to obtain an exact expression for the bulk-to-bulk Green's function of a scalar field in 3D quantum de Sitter space.

 As a first step, we need to identify diffeomorphism invariant observables that create and detect scalar particles along the geodesic world line of an observer. Without loss of generality, we place the semi-classical observer trajectory at the center $r=0$ of the static patch\footnote{Here and in the following we set the de Sitter radius equal to 1.}
\bea
ds^2 \! \is \! -({1-r^2}) d\tau^2 +  \frac{dr^2}{1-r^2} + r^2 d\varphi^2 
\eea 

In metric variables, the relevant observables that feature in the definition of the scalar Green's function are the world-line actions of the observer and their detector (with combined mass $M$) and of the scalar particle (with mass $m$)
\bea
\label{obs}
\,\widehat{\!W\!}{}_{M}({\cal C}) \! \is \!  \exp\Bigl( i \int_{\cal C}\! M d\tau\Bigr) 
\qquad \qquad \widehat{\!G}{}_{m}(\tau_1,\tau_0) \spc = \spc  \exp\Bigl( i   \int^{\tau_1}_{\tau_0}\!\! m\, ds\Bigr)
\eea
Here the respective line integrals run over the trajectory ${\cal C}$ of the observer (semi-classically located at $r=0$) and the trajectory of the scalar particle propagating between two points separated by a proper time distance $\tau_1-\tau_0$ along observer world-line ${\cal C}$. 

The task we will give ourselves is to obtain an exact expression for the expectation value of the dressed scalar two point function defined by the product of the world line observables \eqref{obs} in the de Sitter gravity theory \eqref{gravz}
\bea
\label{greenexp}
G_m(\tau_1,\tau_0)  \is  \int\! [dg] \, e^{-I[g]}\;  \,\widehat{\!G}{}_{m}(\tau_1,\tau_0)\cdot \,\widehat{\!W\!}{}_{M}({\cal C}).
\eea  
Note that, since the observable $\,\widehat{\!G}{}_{m}(\tau_1,\tau_0)$ contains an implicit delta function that enforces that the two end-points of the scalar world line are separated by a proper time distance $\tau = \tau_1-\tau_0$ along ${\cal C}$, it injects additional energy into the observer's detector. The two point function \eqref{greenexp} thus includes an integral over intermediate states with energy $M+\omega$, as indicated in figure 2. We will make this integral explicit later on. In addition, we will account for the exact gravitational back reaction of both world line operators. This is possible thanks to the fact that pure 3D de Sitter gravity is exactly soluble.

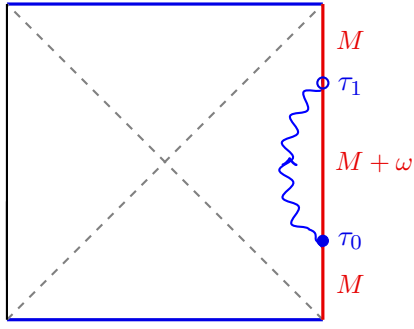
\begin{figure}[t]
 \centering
${}$~~~~~\begin{tikzpicture}[scale=1.05];
\path[draw=blue,thick, snake it] (6,1.05) arc (138:222:1.6cm);
\path[draw=blue,very thick,<-] (5.58,.06) arc (179:181:1cm);
\draw[very thick,\darkblue] (2,-2) -- (6,-2);
\draw[very thick,\darkblue] (2,2) -- (6,2);
\draw[thick,dashed,gray] (2,-2) -- (6,2);
\draw[thick,dashed,gray] (6,-2) -- (2,2);
\draw[thick,\darkred] (2,-2) -- (2,2);
\draw[very thick,\darkredd] (6,-2) -- (6,2);
\draw[thick,\darkred] (2,-1.5) -- (2,1.-1.5);
\filldraw[\darkblue] (6,-1.0) circle (2pt);
\draw[thick, \darkblue] (6,1.0) circle (2pt);
\draw (6.35,-1) node {\rotatebox{0}{\textcolor{\darkblue}{
${\tau_0}$}}};
\draw (6.35,.96) node {\rotatebox{0}{\textcolor{\darkblue}{
${\tau_1}$}}};
\draw (6.35,-1.55) node {\rotatebox{0}{\textcolor{\darkredd}{\small
${M}$}}};
\draw (6.35,1.55) node {\rotatebox{0}{\textcolor{\darkredd}{\small
${M}$}}};
\draw (6.65,0) node {\rotatebox{0}{\textcolor{\darkredd}{\small
${M+\omega}$}}};
\end{tikzpicture}
\vspace{-1mm}
\caption{Green function of a scalar field attached to the worldline  at the center of the static patch.}
\end{figure}

 \usetikzlibrary{patterns}
\subsection{3D Virasoro-de Sitter TFT}

\vspace{-1mm}

The isometry group of 3D de Sitter space is $SL(2,\mathbb{C})$. To quantize pure Einstein gravity on dS$_3$, it is useful to transition to a first order formalism, and rewrite the metric in terms of a dreibein $e^a$ and $SO(2,1)$ spin connection $\omega^a$.  The dreibein and spin-connection naturally combine into an $SL(2,\mathbb{C})$ gauge field $(\Aplus,\Aminus) = (\Aplus^a\spc\ttau_a,  \Aminus^a \ttau_a)$ \cite{Witten:1989ip}. The 3D Einstein-de Sitter action then takes the form of sum of two Chern-Simons actions with imaginary coupling constant $k = i\kappa$ 
\bea
\label{csaction}
S\!  \is \! \frac{i\kappa}{4\pi} \! \int\! {\rm Tr} \Bigl(\Aplus d\Aplus + \frac 2 3 \Aplus\nspc\wedge\nspc\Aplus\nspc\wedge\nspc\Aplus 
\Bigr) - \frac{i\kappa}{4\pi} \! \int\! {\rm Tr} \Bigl( \Aminus d\Aminus + \frac 2 3 \Aminus\nspc\wedge\nspc\Aminus\nspc\wedge\nspc\Aminus 
\Bigr)^{\strut} \nonumber \\[-1mm]\\[-1mm]\nonumber
& & \kappa \, = \, \frac{1}{4G_N}, \qquad \ \Aplus^a \, = \, \omega^a + i e^a, \qquad \Aminus^a \, = \, \omega^a - i e^a
\eea 
Chern-Simons theory with $SL(2,\mathbb{C})$ gauge symmetry has been well studied over the past two decades due to its application to quantum hyperbolic geometry, topological invariants, and the volume conjecture, see e.g. \cite{Dimofte:2011gm}. Here, our interest is to use it as a tool for establishing a correspondence between 3D de Sitter gravity and the double scaled SYK model.

The CS action \eqref{csaction} defines a topological QFT. Depending on the reality conditions and choice of integration contour, it can represent Einstein gravity in dS$_3$, euclidean AdS$_3$, or $S^3$. In all cases, the CS equations of motion
\bea
\label{flat}
F(\Aplus) \! \is \! F(\Aminus) \spc = \spc 0
\eea
imply the constant curvature equations for the metric, along with the torsion constraints that help solve for the spin connection in terms of the dreibein field $e_a$. 

It will often be useful to work in euclidean signature and consider the 3D de Sitter space time as an analytic continuation of the three sphere  $X_1^2 + X_2^2 + X^2_3 + X_4^2 =\, 1$. Introducing coordinates $(X_1,X_2,X_3,X_4) =(\cos\rho \sin \tau, \cos\rho \cos\tau,  \sin\rho \sin \varphi, \sin\rho \cos\varphi)$, 
the round $S^3$ metric takes the form
\bea
ds^2 \! \is \! d\rho^2 + \cos^2\nspc \rho\spc d\tau^2 + \sin^2\nspc \rho \spc d\varphi^2.
\eea
The euclidean representation of the observer trajectory traverses the full thermal circle $\tau \in [0, 2\pi\ra$ at $\rho=0$ and the euclidean observer horizon spans the full spatial circle $\varphi \in [0, 2\pi\ra$ at $\rho = \pi/2$, as indicated in figure 3.

\renewcommand\footnotesize{\small}
The diffeomorphism invariance of the 3D de Sitter gravity theory \eqref{gravz} is replaced by $SL(2,\mathbb{C})$ gauge invariance and topological invariance of the Chern-Simons action. Correspondingly, the gravitational world line observables  \eqref{obs} are represented by suitably designed gauge invariant Wilson lines. 
We will choose the Wilson line of the observer to be of the form\footnote{ Demonstrating the equivalence between the Wilson line observables \eqref{wilsonmo}-\eqref{wilsonmt} and the world line observable \eqref{obs} is a subtle exercise that goes beyond the scope of the present discussion.  However, in the following sections we will present several pieces of evidence that support the proposed identification. For a more detailed treatment, we refer to \cite{Castro:2023bvo, Castro:2023dxp}  and Appendix E in e.g. \cite{Iliesiu:2019xuh}.} 
\bea
\label{wilsonmo}
\,\widehat{\!W\!}{}_{j}({\cal C})  \is  
 \Tr_{{\cal R}_j}\!\nspc\left(P \exp\Bigl(- \oint_{\cal C} \Aplus\Bigr) P \exp\Bigl(- \oint_{\cal C} \Aminus\Bigr)\right)
\eea
where ${\cal R}_j$ denotes a continuous series representation of $SL(2,\mathbb{C})$ with spin $j\! =\! 1/2 \nspc +\nspc is$ and second casimir $c_2({\cal R}_j) = j(1\nspc -\nspc j)  = {M_j^2}/4$. Here ${\cal C}$ denotes the closed contour $\rho=0$. We can also introduce a Wilson lines that goes around the equator contour $\rho = \pi/2$. This would be an observable that measures the mass of the observer at $\rho=0$.

Similarly, we identify the world line action of the scalar particle that travels between points $\tau_1$ and $\tau_2$ along the observer worldline ${\cal C}$ with the open Wilson line 
\bea
\label{wilsonmt}
\,\widehat{\!G}{}_{\Delta}(\tau_1,\tau_0)\! \is \!   P \exp\Bigl(- \int^{\tau_1}_{\tau_0}\!\! \Aplus\Bigr)\, P \exp\Bigl(- \int^{\tau_1}_{\tau_0}\!\! \Aminus\Bigr)
\eea
taken in a representation with second casimir $c_2(\Delta) 
=  \Delta(1\nspc -\nspc \Delta)  = {m_\ddelta^2}/{4}$
and end-points anchored to the observer Wilson line \eqref{wilsonmo} via a suitable $SL(2,\mathbb{C})$ invariant tensor.
The relation between mass parameters $M_j$ and $m_\ddelta$ and $SL(2,\mathbb{C})$ casimirs  is determined via the fact that solutions to the scalar wave equation in $dS_3$  transform as matrix elements of $SL(2,\mathbb{C})$.

\subsection{Quantization}

The first order formalism of 3D de Sitter gravity is well suited for transitioning to the quantum theory and for establishing a correspondence with 2D Virasoro-Liouville CFT with complex central charge. Here we summarize the main steps.

To quantize  3D de Sitter gravity and set up the dS$_3$/CFT$_2$ map, we divide up the three sphere $\MM{}\spc= S^3$ into two half three spheres $\MM{}_{\rm ket}$ and $\MM{}_{\rm bra}$ separated by a 2D cross section~$\Sigma$. Performing the 3D de Sitter gravity functional integral over each half three-sphere with specified boundary conditions on $\Sigma$ will produce corresponding wave functionals $|\Psi_{\rm ket} \ra$ and $\la \Psi_{\rm bra}|$. We can then define the correlation functions by acting with the physical Wilson line observables \eqref{wilsonmo} and \eqref{wilsonmt} on $|\Psi_{\rm ket} \ra$ and taking the inner product with $\la \Psi_{\rm bra}|$. 


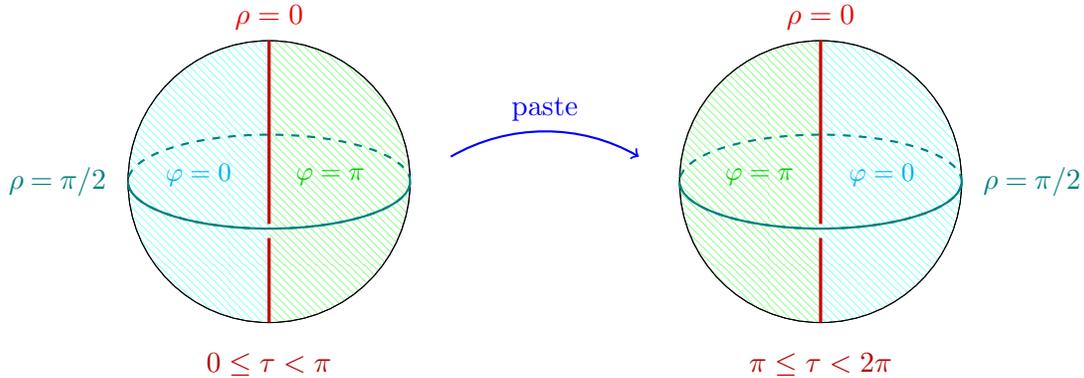
\begin{figure}[t]
\centering
 \begin{tikzpicture}[rotate=0]
\begin{scope}[yscale=1.25,xscale=1.25]
 \tikzset{
    partial ellipse/.style args={#1:#2:#3}{
        insert path={+ (#1:#3) arc (#1:#2:#3)}
    }
}       \draw[thin] (-2,0) circle (1.5cm);
        \draw[pattern={north west lines}, pattern color={cyan!40!white}] (-2,0) [partial ellipse=90:270:1.5cm and 1.5cm];
        \draw[pattern={north west lines}, pattern color={green!40!white}] (-2,0) [partial ellipse=-90:90:1.5cm and 1.5cm];
        \draw[green!50!blue, dashed, thick] (-2, 0) ellipse (1.5cm and 0.5cm);
        \node[green!50!blue] at (-4.25, 0) {$\rho\nspc =\nspc \pi/2$};
        \node[red!90!black] at (-2, 1.75) {$\rho=0$};
        \node[red!70!black]  at (-2, -1.95) {$0 \leq \tau < \pi$};
        \node[green!80!black] at (-1.35, .05) {\small $\varphi =\pi$};
        \node[cyan] at (-2.75, .07) {\small $\varphi =0$};
        \draw[thick, green!50!blue] (-2,0) [partial ellipse=180:360:1.5cm and .5cm];
        \draw[very thick, red!80!black] (-2.00, -1.5) -- (-2.00, -0.6);
        \draw[very thick, red!80!black] (-2.00, -0.45) -- (-2.00, 1.5);
        \end{scope}
\end{tikzpicture}~~~
 \begin{tikzpicture}[baseline={([yshift=-3.5cm]current bounding box.center)}]
        \draw[<-, thick,blue] (60:1) arc (60:120:2.5) node[midway, above] {paste}  ;
\end{tikzpicture}~~~
 \begin{tikzpicture}[rotate=0]
\begin{scope}[yscale=1.25,xscale=1.25]
 \tikzset{
    partial ellipse/.style args={#1:#2:#3}{
        insert path={+ (#1:#3) arc (#1:#2:#3)}
    }
}       \draw[thin] (-2,0) circle (1.5cm);
        \draw[pattern={north west lines}, pattern color={green!40!white}] (-2,0) [partial ellipse=90:270:1.5cm and 1.5cm];
        \draw[pattern={north west lines}, pattern color={cyan!40!white}] (-2,0) [partial ellipse=-90:90:1.5cm and 1.5cm];
        \draw[green!50!blue, dashed, thick] (-2, 0) ellipse (1.5cm and 0.5cm);
        \node[green!50!blue] at (.25, 0) {$\rho\nspc =\nspc \pi/2$};
        \node[red!90!black] at (-2, 1.75) {$\rho=0$};   
        \node[red!70!black]  at (-2, -1.95) {$\pi \leq \tau < 2\pi$};
        \node[cyan] at (-1.35, .07) {\small $\varphi =0$};
        \node[green!80!black] at (-2.65, .05) {\small $\varphi =\pi$};
        \draw[thick, green!50!blue] (-2,0) [partial ellipse=180:360:1.5cm and .5cm];
        \draw[very thick, red!80!black] (-2.00, -1.5) -- (-2.00, -0.6);
        \draw[very thick, red!80!black] (-2.00, -0.45) -- (-2.00, 1.5);
        \end{scope}
\end{tikzpicture}
\caption{Euclidean 3D de Sitter space is obtained by gluing two half three spheres, c.f. \cite{Hikida:2022ltr}. The observer world line at $\rho=0$ (red) and observer horizon at $\rho=\pi/2$ (green) indicated. }
\end{figure}
\vspace{-2mm}

\noindent
The most natural choice for our purpose is to divide up the three sphere into a two equal halves
\bea
\MM{}_{\rm ket} \!\! \is\! \bigl\{\, 0< \varphi< \pi\; \bigr\}, \qquad \qquad
\MM{}_{\rm bra} = \spc \bigl\{\, \pi< \varphi< 2\pi\,\bigr\}.
\eea
Theses two half three-spheres have a common 2D boundary
\bea
\partial \MM{}_{\rm ket} \! \is\!\! - \partial \MM{}_{\rm bra} \, = \,  \bigl\{ \, \varphi = 0, \pi\, \bigr\}  \, \equiv \; \Sigma_{0} \, \cup \, \Sigma_{\pi} 
\eea
in the form of two disks (half two-spheres) $\Sigma_{0}$ and $\Sigma_{\pi}$ glued together along the observer world-line 
\bea
\partial \Sigma_0 \! \is \! - \partial \Sigma_\pi \, = \, \{\, \rho = 0\,\}
\eea
indicated with the red straight line(s) in figure 3.

To quantize the theory, we could either first impose the constraints \eqref{flat} and then quantize, or first quantize and then impose the constraints. The former approach is described in a companion paper \cite{Verlinde:2024znh}. Here we will follow the latter approach.  We promote the $SL(2,\mathbb{C})$ gauge field $\Aplus = (\Aplus,\Aminus)$ into operators that satisfy the commutator algebra 
\bea
\label{acomm}
[\Aplus_{a\alpha}(x), A^+_{b\beta}(y)] \!\is \! -
[\Aminus_{a\alpha}(x), \Aminus_{b\beta}(y)] \spc = \spc 
 \mbox{\Large $\frac{4\pi{\tiny\strut}} k$}\,\delta^{ab} \epsilon_{\alpha\beta}\delta(x-y).
\eea
Physical Hilbert states are represented as wave functionals of a commuting subset of gauge field components. The physical state conditions 
\bea
\label{flatness}
F(\Aplus) \bigl|\Psi \bigr\ra \! \is 
F(\Aminus) \bigl|\Psi\bigr\ra \spc = \spc 0
\eea
then become functional differential equations that uniquely specify the local dependence of the wave functionals on the gauge fields, except for their dependence on the global holonomies around potential operator insertions or non-trivial cycles of the surface $\Sigma$. 

Our first order action \eqref{csaction}  has a non-compact gauge group and imaginary level $k=i\kappa$. Moreover, it is appropriate to look for a quantization procedure that reflects that we are dealing with a 3D gravitational theory. Following the treatment of \cite{Verlinde:1989ua} we decompose
\bea
\label{param}
{\Aplus}=\spc {e}\spc \ttau_+ +  {\omega}\spc \ttau_0 + {f}\spc \ttau_- \qquad\qquad 
{\Aminus} = \spc \bar{f}\spc \ttau_+ + \bar{\omega}\spc \ttau_0 + \bar{e}\spc\spc  \ttau_-.
 \eea
 The $SL(2,\mathbb{C})$ flatness equations \eqref{flatness} then take the form
 \bea
 \label{flatness-two}
 d\smpc{e}+\spc \omega \wedge  {e} =\spc 0, \  & & \quad 
 d\smpc{f} \!-\spc \omega \wedge f =\spc 0, \qquad \quad d\smpc\omega + {e}\wedge f=\spc 0,\nonumber\\[-2.5mm]\\[-2.5mm]\nonumber
 d\bar{f}+\spc \bar\omega \wedge \bar{f} = \spc 0, \  & & \quad 
 \, d\bar{e}-\spc\bar\omega \wedge \bar{e} = \spc 0, \qquad \quad d\smpc\bar{\omega} + \bar{f}\wedge  \bar{e}=\spc 0.
 \eea
and the commutations relations \eqref{acomm} decompose into
 \bea
 [e_\alpha(x),f_\beta(y)] \is\nspc  [\bar{e}_\alpha(x),\bar{f}_\beta(y)]\, =\smpc  \mbox{\Large $\frac{4\pi{\tiny\strut}} k$}\spc \epsilon_{\alpha\beta}\spc\delta(x\nspc-\nspc y)
 \nonumber\\[-2mm]\\[-2mm]\nonumber
  [\omega_\alpha(x),\omega_\beta(y)] \!\is\!  [\bar{\omega}_\alpha(x),\bar{\omega}_\beta(y)] \spc =\spc \mbox{\Large $\frac{4\pi{\tiny\strut}}{k}$}\spc \epsilon_{\alpha\beta}\spc\delta(x\nspc-\nspc y).
 \eea
 Note that the flatness conditions \eqref{flatness-two} look like the torsion and constant curvature constraints for two 2D metrics parametrized by the zweibeins and spin connections $(e,f,\omega)$ and $(\bar{e},\bar{f},\bar\omega)$. 

We need to select which commuting subset of variables to let our wave functionals depend on. In terms of the path integral, this  amounts to picking suitable Dirichlet boundary conditions on the quantization surface $\Sigma$. Let us split $\omega = \omega_z dz +\omega_{\bar z} d\bar{z}$ and $\bar\omega = \bar\omega_z dz +\bar\omega_{\bar z} d\bar{z}$. Following  \cite{Verlinde:1989ua} we let wave functions depend on $({e},\omega_z)$ and  $(\bar{e},\bar\omega_{\bar z})$. In \cite{Verlinde:1989ua} it was shown that the physical state conditions \eqref{flatness} then reduce to the Virasoro Ward identities of a pair of chiral 2D CFTs. We will not present the complete analysis here, but only summarize the main result. For a more detailed discussion, see \cite{Verlinde:1989ua} \cite{Cotler:2019nbi,Collier:2023fwi,Collier:2024mgv}).

We denote the ket state of the 3D de Sitter gravity theory by $|\Psi_+\ra$ and the bra state by $\la \Psi_-|$. The physical meaning of the $+$ and $-$ subscripts will become clear shortly. In view the split form of the first order gravity action \eqref{csaction}, we (temporarily) make the Ansatz that the  wave function factorizes into the product of two chiral wave functions\footnote{Here $\omega = \omega_z$ and $\bar{\omega}= \bar{\omega}_{\bar z}$ denote only the  chiral component of the spin connection one-forms.}
\bea
\label{psidef}
\la e,\bar{e},\omega,\bar{\omega} |\Psi \ra \is   \la e,\omega|\spc \psi_+ \ra \la \bar{e},\bar{\omega}|\,\overline{\nspc\psi\nspc}_- \ra 
\eea
We further introduce the parametrization $e= e^{\sigma}(dz + \mu d\bar{z})$ and $\bar{e} = e^{\bar{\sigma}}(d\bar{z} + \bar\mu d{z})$.
The flatness conditions \eqref{flatness-two} can then be explicitly solved in terms of (chiral) two-dimensional conformal field theory partition functions as follows
\bea
\la e,\omega| \psi_+\ra\!\! \is\!\nspc e^{-S_+(\sigma,\omega,\mu)}\smpc \la\smpc \mu\smpc |\smpc \psi_+\ra,
\nonumber\\[-1.75mm]\label{confanom} \\[-1.75mm]\nonumber 
\la \bar{e},\bar{\omega}| \psi_-\ra \! \is \! 
e^{-S_-(\bar\sigma,\bar\omega,\bar\mu)}\smpc \la\smpc \bar\mu \smpc | \smpc \psi_-\ra.{}_{\strut}
\eea
Here $S_+(\sigma,\omega,\mu)$ and $S_-(\bar\sigma,\bar\omega,\bar\mu)$ are given by a chiral Liouville action  \cite{Verlinde:1989ua}, each with opposite chirality. Defining 
\bea
 \la \smpc T \smpc \ra \equiv \frac{\delta}{\delta\mu}\log \la\smpc\mu\smpc |\smpc \psi_+\ra \  & &  \ \la \smpc \bar{T} \smpc \ra \equiv \frac{\delta}{\delta\bar\mu}\log \la\smpc\bar\mu\smpc |\smpc \psi_-\ra
 \eea
the remaining flatness equations take the familiar form 
of the Virasoro Ward identities 
\bea
\bigl(\overline\partial\nspc -\nspc \mu\partial\nspc -\nspc 2\partial\mu\bigr)\la\spc T\spc \ra\!\!  \is\!\nspc - \frac{\raisebox{-1pt}{$c_+$}{}_{\tiny\strut}\!\!}{\raisebox{.5pt}{\small 12}}\,\spc\partial^3 \mu , 
\nonumber\\[-1.75mm]\label{virasoro-wi} \\[-1.75mm]\nonumber
\bigl(\partial\nspc - \nspc\bar\mu\bar\partial\nspc -\nspc 2\bar\partial\bar\mu\bigr) \la\spc \bar{T}\spc\ra 
\!  \is\!
 - \frac{\raisebox{-1pt}{$c_-$}{}_{\tiny\strut}\!\!}{\raisebox{.5pt}{\small 12}}\,\spc \bar{\partial}{}^3 \bar\mu 
\eea 
of the left- and right-moving stress energy tensor of two chiral 2D CFTs with central charges\footnote{Recall that $\kappa=1/4G_N$. Equation \eqref{centralp} matches with the familiar AdS$_3$/CFT$_2$ relation $c = 3/2G_N$, except for the factor of $\pm i$. The shift by 13 is a one-loop correction \cite{Cotler:2019nbi}; pure dS$_3$ gravity is one-loop exact. }
\bea
\label{centralp}
c_\pm \!\is\nspc 13 \pm 6i\bigl(\kappa - \kappa^{-1}\bigr) 
\eea
Intriguingly,  the two central charges add up to the critical value $c_+ + c_- = 26$. As we will see shortly, this fact provides some useful guidance for how to define the inner product on the physical Hilbert space of the 3D de Sitter gravity theory.

Solutions to the Virasoro Ward identities \eqref{virasoro-wi} are, essentially by definition, equal to the conformal blocks of a universal Virasoro CFT with central charge $c_+$ and $c_-$. The  physical 3D de Sitter gravity wave functions $\la\smpc \mu \smpc | \smpc \psi_+\ra$ and  $\la\smpc \bar\mu \smpc | \smpc \psi_-\ra$ can thus expressed as chiral CFT partition functions in the presence of a non-zero Beltrami deformation of the metric $\la\smpc \mu \smpc |\spc \psi_+\ra =\bigl\la \exp\bigl( {-\mbox{$\int$} \mu\spc T\smpc  } \bigr) \bigr\ra_{\! +} $ and $\la\smpc \bar{\mu} \smpc |\spc\psi_-\ra \spc = \spc  \bigl\la \exp\bigl( {-\mbox{$\int$}\bar{\mu}\spc \bar{T}\smpc  } \bigr)\bigr\ra_{\! -}$. The exponential prefactors for the full wave functions  \eqref{confanom} are functionals that represent the Weyl anomaly of the chiral 2D CFTs \cite{Verlinde:1989ua}.

Next let us discuss the wave function representation of the dual Hilbert states. We choose our reality conditions such that $(e,\omega)$ and $(\bar{e},\bar{\omega})$ are interchanged under complex conjugation. Our wave functions are thus defined with a real polarization. This means the dual Hilbert states are also functions of the same variables. Repeating the above analysis for the bra state $\la \bar\Psi| = \la\bar{\psi}_+|\la{\psi}_-|$ leads to an identification of the wave functions with the chiral Virasoro CFT partition function with the same central charge \eqref{centralp} but with opposite chirality. Combining the bra and ket states, we thus have four chiral Virasoro CFT partition functions. It will be useful to (for now, schematically) represent them as the functional integral of four chiral Liouville field theories 
\bea
\la \smpc{e}\smpc |\psi_+ \ra\!\is \! \int[d \phi_+] \, e^{- S^+_L(\phi_+,{e})}\bigr|_{\rm chiral} \qquad \qquad \la \bar{\psi}_+| \smpc\bar{e}\smpc \ra \spc = \spc  \int[d \phi_+] \, e^{- S^+_L(\phi_+,\bar{e})}\bigr|_{\rm \overline{chiral}}
\nonumber\\[-1.75mm]\label{virasoro-wit} \\[-1.75mm]\nonumber
  \la \smpc\bar{e}\smpc|\spc{\psi}_-\ra \!\is\! \int[d \bar\phi_-] \, e^{-S^-_L(\bar{\phi}_-,\bar{e})}\bigr|_{\rm \overline{chiral}}
  \qquad \qquad  \la \bar{\psi}_-| \smpc{e}\smpc\ra \spc = \spc  \int[d \phi_-] \, e^{- S^-_L(\phi_-,{e})}\bigr|_{\rm {chiral}}\nonumber
\eea
Here, for notational simplicity, we suppressed the dependence on the spin connections $\omega$ and $\bar{\omega}$.

\subsection{Inner product}
\vspace{-1.5mm}

We now make the natural step of combining the chiral partition functions into non-chiral partition functions. Concretely,
we can transition to a metric formulation by introducing a 2D metric $g_{\alpha\beta} = \eta_{ab} e^a_\alpha \bar{e}{}^b_\beta$ and eliminating the spin connection variables via the torsion constraints.\footnote{As shown in \cite{Verlinde:1989ua}, the $\omega$-dependence of the physical wave functionals takes the form of a simple gaussian pre-factor. In physical expectation values, one can perform the gaussian $\omega$ integral, resulting in a fully covariant 2D expression for inner product between the bra and ket functionals. The saddle point equation equals  the torsion constraint. For simplicity of presentation, we bypass this step here.}
The product $ \la \bar{\psi}_+| \smpc\bar{e}\smpc \ra\la \smpc{e}\smpc |\psi_+ \ra$ and  $\la \bar{\psi}_-| \smpc{e}\smpc\ra \spc  \la \smpc\bar{e}\smpc|\spc{\psi}_-\ra$ of the chiral ket and bra wave functions then assemble into covariant functionals of the metric $g_{\alpha\beta}$ satisfying the same conformal Ward identities as the non-chiral partition functions of  Virasoro CFTs with central charge $c_+$ and $c_-$ 
\cite{Verlinde:1989ua}. We represent these partition functions by means of a functional integral of two space-like Liouville CFTs 
\bea
\label{partone}
 \la \bar{\psi}_+| \smpc\bar{e}\smpc \ra\la \smpc{e}\smpc |\psi_+ \ra
\!\nspc \is \!\nspc \int[d \phi_+] \, e^{- S^+_L(\phi_+,\smpc g\smpc)} \equiv  \la\spc g\spc | \Psi_+\ra\nonumber\\[-2mm]\\[-2mm]\nonumber
 \la \bar{\psi}_-| \smpc{e}\smpc\ra \spc  \la \smpc\bar{e}\smpc|\spc{\psi}_-\ra\!\nspc \is \!\nspc \int[d \phi_-] \, e^{- S^-_L(\phi_-,\smpc {g}\smpc)}\equiv \la\Psi_-| \spc{g} \spc \ra
\eea
With these identifications, we trade the ket and bra states $|\Psi\ra$ and  $\la\bar\Psi|$, defined in \eqref{psidef} as the path integral over the half three spheres $\MM_{\rm ket}$ and $\MM_{\rm bra}$, for the new wave functions $|\Psi_+\ra$ and $\la \Psi_-|$. The full partition function of the 3D de Sitter gravity can be written as an inner product obtained by integrating the product of the two CFT partition functions \eqref{partone} over all 2D metrics\footnote{The equivalence between the two descriptions follows from the schematic rearrangement
$$\la\,\overline{\!\Psi\!}\, | \Psi\ra\nspc =\! \int\! [d\smpc e d\bar{e}]\spc\la \,\overline{\!\Psi\!}\,  | {e}, \bar{e}\ra \la{e}, \bar{e}\smpc |\Psi\ra=\!
 \int\! [d\smpc e d\bar{e}]\spc \la \bar{\psi}_+| \bar{e} \ra\la \smpc{e}\smpc |\psi_+ \ra\la \bar{\psi}_-| {e} \ra   \la \bar{e}|{\psi}_-\ra \equiv
\int [dg] \la\Psi_- | g\ra \la g | \Psi_+\ra \spc = \spc \la \Psi_- | \Psi_+\ra $$}. 
The total Weyl anomaly in the integrand adds up to $c_++c_-=26$. 
\bea
\la\spc \Psi_-|\spc \Psi_+ \ra\!\nspc
 \is  \int[dg] \int[d \phi_+d\phi_-] \, e^{-S^+_L(\phi_+,\smpc g\smpc)-S^-_L(\phi_-,\smpc g\smpc)}
\eea

A priori, both partition functions \eqref{partone} are defined to live on the 2D surface $\Sigma =\Sigma_0 \cup \Sigma_\pi$. To make the correspondence with the SYK model, we now impose the $\mathbb{Z}_2$ invariance condition that the theory is gauge invariant under the  inversion (here the superscript on $\Aminus^T$ denotes transposition)
\bea
\label{cpt}
(\spc \tau,\spc \varphi, \spc \Aplus,\Aminus^T\spc )\spc{\small\strut}  &\sim&  (\spc \tau+\pi\nspc, \spc \pi -\varphi\nspc, \spc \Aminus^T,\Aplus \spc).
\eea
This transformation flips the sign of all embedding coordinates except $X_3$ and thus breaks the $SO(4)$ symmetry to $SO(3)$, or after Wick rotation to lorentzian signature, $SL(2,\mathbb{C})$ to $SU(1,1)$, c.f. \cite{Verlinde:2024znh}.
The transformation \eqref{cpt} interchanges the two half spheres $\Sigma_0$ and $\Sigma_\pi$, while mapping $\Aplus$ to $\Aminus^T$. Hence we can use this identification to restrict the wave functions to live only on one of the two half spheres, say $\Sigma_0$. In terms of the 2D Liouville path-integral, taking the $\mathbb{Z}_2$-quotient amounts to imposing reflecting (FZZT) boundary conditions at the observer worldline at $\rho=0$.

Gravitational correlation functions are obtained by acting with physical operators $\,\widehat{\!W\!}$ on the physical state $|\Psi_+\ra$ and then taking the inner product with $\la \Psi_-|$ \cite{Verlinde:1989ua, Collier:2023fwi, Collier:2023cyw}. The inner product between two physical states is defined by multiplying them together to produce a disk partition function or correlation function of the combined $c = 26$ theory, factor in a $c=-26$ $(b,c)$ ghost partition function, and integrate the result over all moduli introduced by  the physical operator $\,\widehat{\!W\!}$
\bea
\bigl\langle  \Psi_{+}\spc \bigl|\, \widehat{\!W} \bigr| \spc {\Psi_-}\bigr\rangle \! \is \! \int_{\cal M} \,  Z_{\rm gh}\, 
\Psi_{-}\, \widehat{\!W} \, {\Psi_+} 
\eea
This inner product is naturally defined operation thanks to the special property that the total central charge of the two LCFTs and ghost fields vanishes.

At this point, it is fitting to make the transition from the 3D gravity formalism in terms of $SL(2,\mathbb{C})$ gauge fields and Wilson lines to the well studied world of 2D CFTs, albeit in the slightly unfamiliar setting with complex central charge. Here we indicate the proposed 3D/2D dictionary in a few words, leaving a more detailed study until after we introduce the relevant 2D CFTs.

As a first correspondence, we propose that there exists a one-to-one map between closed 3D Wilson lines $\,\widehat{\!W\!}({\cal C})$ acting along ${\cal C} =\partial \Sigma_0$ and 2D Verlinde loops acting on FZZT branes in the 2D Virasoro CFT \cite{Gaiotto:2014lma}. As we will see in the next few sections, Virasoro-Liouville CFTs with complex central charge possess a preferred FZZT brane $|s_0\ra$ with zero boundary cosmological constant $\mu_B$. We postulate that the corresponding disk partition function represents the state $|\Psi_+\ra$. Acting with 3D Wilson lines on this state produces other FZZT boundary states $|s_1\ra = |s_0 + \omega\ra$ with non-zero energy relative to~$|s_0\ra$, as shown in figure 4. In a separate paper \cite{Verlinde:2024znh}, we study these loop operators in more detail from the point of view of the constrain first and quantize second approach.

As a second entry of the dictionary, we propose that the open Wilson line $\,\widehat{\!G}_\Delta(\tau_1,\tau_0)$ amounts to acting with two boundary condition changing operators $V_\Delta(\tau_0)$ and $V_\Delta(\tau_1)$ on the FZZT state. The two operator insertions are accompanied by a delta function that enforces that the two operators are separated by a proper time distance $\tau_1-\tau_2$ as measured by the dynamical boundary metric. In the following two sections, we will compute the corresponding two point function by borrowing known results obained in the study of 2D Liouville CFTs with complex central charge \cite{Zamolodchikov:2005fy}. We will then compare the result with the two point function of the SYK model.

\begin{figure}[t]
\centering
 \begin{tikzpicture}[rotate=0]
\begin{scope}[yscale=1.1,xscale=1.1]
 \tikzset{
    partial ellipse/.style args={#1:#2:#3}{
        insert path={+ (#1:#3) arc (#1:#2:#3)}
    }
}       
        \draw[pattern={north west lines}, pattern color={cyan}] (-2,0) circle (1.5cm);
        \node[red!70!black] at (-2, -1.2) {$\rho=0$}; 
        \node[cyan!80!white] at (-2, 0) {\large $\Sigma_0$};
 \draw[very thick, red!80!black] (-2,0) circle (1.5cm);  
   \node[red!60!black] at (-3.75, 0) {$s_0$};  
        \end{scope}
\end{tikzpicture}~~~~~~~~~~~~~~~~~~~~~~~\begin{tikzpicture}[rotate=0,baseline={([yshift=-1.55cm]current bounding box.center)}]
\begin{scope}[yscale=1.1,xscale=1.1]
 \tikzset{
    partial ellipse/.style args={#1:#2:#3}{
        insert path={+ (#1:#3) arc (#1:#2:#3)}
    }
}       
        \draw[pattern={north west lines}, pattern color={cyan}] (-2,0) circle (1.5cm);
        \node[blue!90!black] at (-.95, -1.49) {$\tau_0$};  
        \node[blue!90!black] at (-.95, 1.45) {$\tau_1$};
        \node[cyan!80!white] at (-2.03, 0) {\large $\Sigma_0$};
        \node[blue!90!black] at (-1.1, 0) {$\Delta$};
 \draw[very thick, red!80!black] (-2,0) circle (1.5cm);  
 \path[draw=blue,thick, snake it] (-1,1.12) arc (145:215:1.96cm);  
 \draw[very thick, blue!80!black] (-1,-1.1) circle (.5mm);
 \draw[very thick, blue!80!black] (-1,1.1) circle (.5mm);
   \node[red!60!black] at (-3.75, 0) {$s_0$};  
        \node[red!60!black] at (-.15, 0) {$s_1$};
        \end{scope}
\end{tikzpicture}
\vspace{-1mm}
\caption{The 3D functional integral over $\MM{}_{\rm ket}$ equals the 2D disk partition function. The  scalar Green's function corresponds to the correlation function of two boundary operators.}
\vspace{-1mm}
\end{figure}
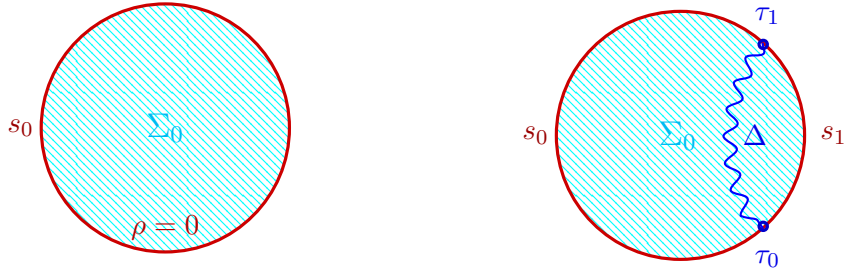

\section{Some Properties of 2D Liouville-de Sitter Gravity}
 
 \vspace{-2mm}
 
Motivated by the above correspondence, we introduce the 2D gravity model specified by the sum $S =S[\phi_+] + S[\phi_-]$ of two space-like Liouville conformal field theories 
\bea
S[\phi_+]\is \frac{1}{4\pi}\, \int_{\! \MM{}\nspc}\Bigl[ \partial \phip\bar\partial\phip  + Q_+ R \, \phi_+ 
+ \mu_+ e^{2b_+ \phip}\Bigr]  \, + \, \frac{1}{2\pi} \int_{\partial M} \!\! \bigl(Q_+ k + \mu_B \spc e^{b_+\phip}\bigr) 
\nonumber \\[3.25mm]
\label{lcftsum}
S[\phi_-] \is \frac{1}{4\pi} \int_{\! \MM{}\nspc}\spc \Bigl[  \partial \phim\bar\partial\phim + Q_-  R \, \phi_- 
+ \mu_-  e^{2b_-\phim}\Bigr] \, + \,  \frac {1}{2\pi}\int_{\partial M}  \!\! \bigl(Q_-k + \mu_B \spc e^{b_-\phim}\bigr) 
\label{bpm}\\[3.5mm]
& & \qquad \ \ \ Q_{{\!\spc}_{\pm}} = \spc b_{{\!\spc}_{\pm}} + {b^{-1}_{{\!\spc}_{{}^{\pm}}}\!\!},\ \ \qquad\  b^{-1}_\pm = \spc e^{\pm i \pi /4}\beta^{-1}\nonumber
\eea 
We will consider the theory on the 2D disk $\Sigma_0$ with FZZT boundary conditions specified by the boundary cosmological constant $\mu_B$. In correlators, $\mu_B$ will be piece-wise constant.
The background charge $Q_\pm$ improves the CFT energy-momentum tensor to ensure that the cosmological constant terms $e^{2b_\pm \phi_\pm}$ and $e^{b_+\varphi}$  transform respectively as (1,1) forms and as 1-forms. Note that we have chosen $b_\pm$ to be both complex. Upon quantization, the 2D gravity theory describes a sum of two conformal field theories with complex central charges  
\bea
c_\pm \!\is \! 1 + 6 Q_\pm^2 = \, 13 \pm 6i \bigl(\beta^{-2} - \beta^{2}\bigr)\qquad \quad
c_+ + c_- = 26. 
\eea 
Just as in critical string theory, we can gauge the combined infinite dimensional Virasoro symmetry of the two spacelike LCFTs by adding $(b,c)$ ghosts and requiring BRST invariance. 
We will call the 2D theory defined by this procedure Liouville-de Sitter gravity.

We will be interested in physical boundary operators in Liouville-de Sitter gravity that satisfy the BRST condition $[Q,{\cal O}]=0$. These operators are given by the integral over the disk boundary $\partial \Sigma_0$ of local vertex operators with total conformal dimension~1
\bea
\label{vphys}
\qquad {\cal O}_{\Delta}^{\spc \rm phys} \! \is\! \int
  \textcolor{black}{V^+_{\Delta}} \,  \textcolor{\darkred}{V^-_{1 - \Delta}}
  \qquad \quad \begin{array}{cc}{ \textcolor{black}\,{V^+_{\Delta}\, \, = \, e^{\alpha_+ \phi_+}}} 
  & {\textcolor{black}{\alpha_{\!\spc {}_+} \! =  b_+\Delta \ \ \ }}
  \\[3.5mm]
 {\textcolor{\darkred}{V^-_{1 - \Delta} = e^{\alpha_- \phi_-}\ }}
&    {\textcolor{\darkred}{{\ \ \alpha_{\!\spc {}_-} \! = \,  b_-(1-  \Delta)}}}
\end{array}
\eea
We can then define the two point function of  two such physical operators via
\bea
\label{oamplitude}
\Bigl\la {\cal O}_{\Delta}^{\spc \rm phys}(\tau_1){\cal O}_{\Delta}^{\spc \rm phys}(\tau_0) \Bigr\ra \is \int^{\strut}_{\strut}\! [d\phi_+d\phi_-] [ db dc] \; e^{iS[\phi_+] + i S[\phi_-]  + S[b,c]}\; {\cal O}_{\Delta}^{\spc \rm phys}(\tau_1){\cal O}_{\Delta}^{\spc \rm phys}(\tau_0)
\eea
Here the arguments $\tau_1$ and $\tau_0$ indicate the restriction that the two operators are separated by a proper time distance $\tau_1-\tau_0$ along the disk boundary (see section 4.2).

In the next two sections we will compute this two point function and compare it with the two-point function of physical operators in the doubled SYK model. An early hint that the two point function might match is that the expression \eqref{gexact} for the SYK two-point function has appeared before in a paper by Zamolodchikov \cite{Zamolodchikov:2005fy} in the study of the analytically continuation of the DOZZ formula of 2D Liouville CFT.

\def\tplus{{\mbox{\tiny $+$}}}
\def\tminus{{\mbox{\tiny $-$}}}
\def\CC{\rmC}

\subsection{Zamolodchikov's Formula}
\vspace{-1.5mm}

For reasons that will become clear soon, we wish to compute the product of the non-chiral  three point functions of two  space-like Liouville CFTs with total central charge $c_++c_-=26$
\bea
\label{product}
  \la \smpc s_{0}\smpc | \smpc V^{+}_\Delta\spc V^{-}_{1-\Delta}\smpc |\smpc s_{1}\smpc \ra \! \is\! \CC^+_{\smpc \Delta}\smpc(s_{0}, s_{1})\,  \CC^-_{\smpc1\nspc\smpc-\nspc\smpc\Delta}\nspc(s_{0}, s_{1}\nspc)
\eea
defined as the matrix element of two Liouville vertex operators with left and right-moving scale dimension $\Delta$ and $1- \Delta$ between two Liouville momentum  eigen states $
|\spc s_0 \spc \ra = | s_0\ra_+ |s_0\ra_-$ and $|\spc s_1 \spc \ra=  | s_1\ra_+ |s_1\ra_-$.
The  quantum numbers $s_i$ label the conformal dimensions of the states~via 
\bea
\Delta_\pm({s}) \! \is \! \frac{Q_\pm^2}{4} + b_\pm^2 s^2 \, = \,  \frac{c_\pm\!-1}{24} \mp i \beta^2 s^2.
\eea
Note that the sum of all conformal dimensions of the $+$ and $-$ states add up to 1, as prescribed by BRST invariance. The three-point function \eqref{product} thus equals the physical three-point function of the critical $c=26$ gravity theory defined by the sum of the two space-like Liouville CFTs.   Its explicit form can be obtained directly by using known LCFT formulas. However, it is instructive to first recall an old result 
\cite{Zamolodchikov:2005fy} which predicts the answer for us.

In a beautiful paper \cite{Zamolodchikov:2005fy}, Zamolodchikov studies the analytic continuation of the DOZZ three point function to arbitrary complex values of the charge $c$. He derives three remarkable results. 
First, he introduces a new class of theories given by a ``timelike" Liouville theory defined for general complex central charge $c_T$, including the range $\Re(c_{\smpc \rm T}) =13$. 
Secondly, he computes the  three-point functions $\CC_\Delta^{\smpc \rm T}(\smpc s_0, s_1)$ of the timelike and shows that they are (upto some elementary leg-pole factors) the inverse of the DOZZ three-point function  $\CC_{1\nspc -\nspc \ddelta}^{-}(\smpc s_0, s_1)$ of the corresponding shadow operators in the spacelike Liouville CFT with complementary central charge $c^-_{\smpc \rm L} = 26 - c_{\smpc \rm T}$.  We write this result schematically as
\bea
\label{prodone}
\CC^{\smpc \rm T}_\ddelta(\smpc s_0, s_1) \spc \CC^-_{1\nspc - \nspc \ddelta} (s_0, s_1) \! \is \! 1.
\eea

A third result reported in \cite{Zamolodchikov:2005fy} concerns the ratio between the timelike Liouville 3-point function $\CC^{\smpc \rm T}_\ddelta(\smpc s_0, s_1)$ and the 3-point function $\CC^+_{\ddelta} (s_0, s_1)$ of a spacelike Liouville CFT with the same central charge $c^+_{\smpc \rm L} = c_{\smpc \rm T}$. We denote this ratio as
\bea
\label{ratio}
{\CC^+_{\ddelta} (s_0, s_1)}/
{\CC^{\smpc \rm T}_\ddelta(\smpc s_0, s_1) }
\eea
Liouville CFT and generalized minimal models, when extended to complex central charge, have a complementary region of validity but the regions overlap along the line $\Re{c} = 13$. 
So the ratio \eqref{ratio} can be defined in our regime of interest.

By inserting the identity \eqref{prodone} into the ratio \eqref{ratio}, we can employ the result of \cite{Zamolodchikov:2005fy} to obtain the product \eqref{product} of LCFT three point functions.
Equation (59) in \cite{Zamolodchikov:2005fy} then gives that 
\bea
\label{magicformula}
 \CC^+_{\smpc \Delta}\smpc(s_{0}, s_{1})\,  \CC^-_{\smpc1\nspc\smpc-\nspc\smpc\Delta}\nspc(s_{0}, s_{1}\nspc)
\! \is  \frac{\spc \vartheta_1  \bigl( 2 i\Delta, \tilde{q}\spc\bigr)
\prod_a {\vartheta_1  \bigl( 2 s_{a}, \tilde{q}\spc\bigr)} 
 {\strut}}{\smpc \vartheta_1  \bigl({\smpc  i\Delta\pm s_0  \pm  s_1}, \tilde{q}\spc\bigr)}{ \strut} 
\eea
with $a\in\{\Delta,0,1\}$, $b\Delta =Q/2 + is_\Delta$ and $\tilde{q}= e^{2\pi i/ b^2}$.
We call this identity Zamolodchikov's formula. 
The expression \eqref{magicformula} has several remarkable properties.  In particular, it is invariant under modular transformations that replace $b$ with $1/b$.  

\begin{figure}[t]
\centering
\begin{tikzpicture}[scale=.25,rotate=0] 
 \foreach \x in {6} {
 \draw[gray, dashed] (1.2 * \x,.5) -- (1.2 * \x,11.5);
}
 \foreach \y in {6} {
 \draw[gray, dashed] (.5,\y) -- (13.8,\y);
}
\foreach \x in {1,2,...,6} {
    \foreach \y in {1,2,...,6} {
      \draw[thick,\darkblue] (1.2 * \x,\y) circle (0.15); 
    }
    }
\foreach \x in {7,8,...,11} {
    \foreach \y in {7,8,...,11} {
      \draw[thick,\darkblue] (1.2 * \x,\y) circle (0.15); 
    }
    }
    \draw (7,-1) node {\textcolor{\darkblue}{$\Upsilon_b(z)$}};
    \end{tikzpicture}~~~~~~~~~~~~~~\begin{tikzpicture}[scale=.25,rotate=0]
  \foreach \x in {6} {
 \draw[gray, dashed] (1.2 * \x,.5) -- (1.2 * \x,11.5);
}
  \foreach \y in {1,...,11} {
    \draw[white,very thin] (.5,\y) -- (13.8,\y);
  }
 \foreach \y in {6} {
 \draw[gray, dashed] (.5,\y) -- (13.8,\y);
} 
\foreach \x in {1,2,...,6} {
    \foreach \y in {1,2,...,6} {
      \draw[thick,\darkblue] (1.2 * \x,\y) circle (0.15); 
    }
    }
    \foreach \x in {7,8,...,11} {
    \foreach \y in {7,8,...,11} {
      \draw[thick,\darkblue] (1.2 * \x,\y) circle (0.15); 
    }
    }
\foreach \y in {1,2,...,6} {
    \foreach \x in {7,8,...,11} {
      \draw[thick,\darkredd] (1.2 * \x,\y) circle (0.15); 
    }
    }
    \foreach \x in {1,2,...,6} {
    \foreach \y in {7,8,...,11} {
      \draw[thick,\darkredd] (1.2 * \x,\y) circle (0.15); 
    }
    }
    \draw (7.3,-1) node {\textcolor{\darkblue}{$\Upsilon_b(z)$}\textcolor{\darkredd}{$\Upsilon_{ib}(-iz+ib)$}};
\end{tikzpicture}~~~~~~~~~~~~~~~~\begin{tikzpicture}[scale=.25,rotate=0]
\foreach \x in {1,2,...,11} {
    \foreach \y in {1,2,...,11} {
      \draw[thick,\darkblue] (1.2 * \x,\y) circle (0.15); 
    }
    }
    \draw (7.5,-1) node {\textcolor{\darkblue}{$\vartheta_1(z,q)$}};
\end{tikzpicture}

\vspace{-1mm}
\caption{A product of two $\Upsilon$-functions has the same regular lattice of zeros as the theta function. }
\vspace{-1mm}
\end{figure}
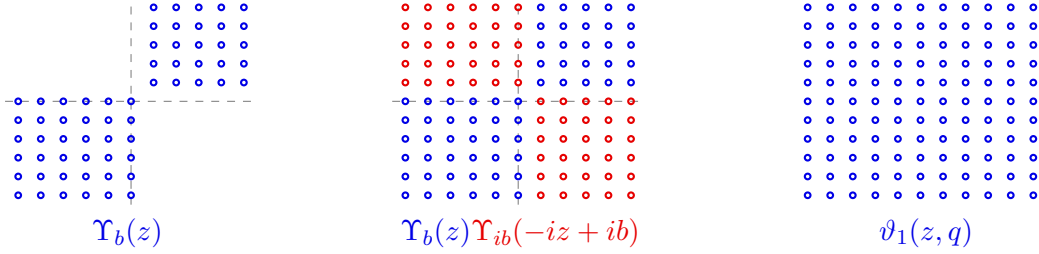

To write the identity that leads to \eqref{magicformula}, it is convenient to denote $b_+ = b$ and $b_- = ib$. 
The DOZZ three point functions of the two Liouville theories are then given by \cite{Zamolodchikov:2005fy} 
\bea
\label{dozzs}
 \CC^+_{\smpc \Delta}\smpc(s_{0}, s_{1})
\nspc \! \is \!\nspc
\frac{ \; \prod_a \Upsilon_{\! b}(-2ibs_a)
}
{\Upsilon_{\nspc b}\bigl(b(\Delta  \pm is_{0}\nspc  \pm\nspc is_1)\bigr)} \qquad \qquad 
\CC^-_{\smpc1\nspc\smpc-\nspc\smpc\Delta}\nspc(s_{0}, s_{1}\nspc)
\smpc =\smpc
\frac{ \; \prod_a \Upsilon_{\nspc ib}\bigl(ib\nspc -\nspc 2bs_a\bigr)^{\strut}}
{\Upsilon_{\nspc ib}\bigl(ib-ib(\Delta \pm is_{0}\nspc  \pm\nspc is_1)\bigr)_{\strut}}\ \
\eea
with $\Upsilon(z)$ the $\Upsilon$-function defined in (see Appendix A).
$ \Upsilon_{\! b}(z)$ is an entire function with an infinite number of zeros located at
$z=-nb-{m}{b}^{-1} $ and $z=(n'\nspc+\nspc 1)b+ ({m'\nspc+\nspc 1}){b^{-1}}$ with $m,m',n,n'\geq 0$, as indicated on the left in figure 5.  The  special identity that leads to  formula \eqref{magicformula} reads
\bea
\Upsilon_{\nspc b}(bz) \Upsilon_{ib}(ib-ibz) \!\is \! e^{i\frac\pi 8 (b-b^{-1} - 2bz)^2-\frac{i\pi}{4b^{2}}}\frac{\vartheta_1( z, \tilde{q})}{\vartheta_3(0,\tilde{q})}
\eea
A proof of this formula is given in \cite{Zamolodchikov:2005fy}.  Geometrically, it follows from the fact that the product of $\Upsilon$-functions on left-hand side has the same regular lattice of zeros as the theta function on the right-hand-side. A pictorial representation of this identity is given in figure 5.

\subsection{Boundary Proper Time}

\vspace{-1mm}

The total LCFT action \eqref{lcftsum} contains two boundary terms proportional to the boundary cosmological constant $\mu_B$. For fixed $\mu_B$, the boundary action implements FZZT boundary conditions. Here we will show that the standard parametrization of $\mu_B$ in terms of Liouville momenta exactly matches with the parametrization of the SYK energy in terms of the spectral angle $\theta = \lambda s$.

For real central charge $c$ and coupling $b$, FZZT boundary states $|s\ra$ labeled by a spectral variable $s$ related to the boundary cosmological constant $\mu_B$ via \cite{Fateev:2000ik, Teschner:2000md}
\bea 
\label{mubrel}
\mu^2_B \! \is\! \frac{2 \mu_L  \cosh^22\pi b^2s}{{|\smpc1\!\smpc -\!\smpc q\smpc |}}, \qquad \quad q=e^{2\pi i b^2}
\eea
Here $\mu_L$ denotes the bulk cosmological constant. This boundary state $|s\ra$ is a Cardy state, defined by property that the CFT on a strip bordered on one side by the FZZT state $|s\ra$ and on the other side by a ZZ state $|ZZ\ra$ is projected, along the longitudinal channel parallel to the two boundaries,  onto the Virasoro module associate with the primary state with conformal dimension $\Delta = \frac{c- 1}{24} + s^2$. 

The parametrization \eqref{mubrel} of the boundary cosmological constant $\mu_B$ in terms of the momentum $s$ strongly resembles the expression \eqref{frulestwo} of the SYK energy. Indeed, it is a well-motivated hypothesis that we should identify the SYK energy with the $\mu_B$ parameter of the 2D gravity theory. Physically, it amounts to adopting the following diffeomorphism invariant definition of the coordinate difference $\tau_1-\tau_2$ between two points $1$ and $2$ along the boundary
\bea
\tau_2-\tau_1 \is \int_{1}^{2} \! d\tau \spc = \spc \frac 1 2 \int_{1}^2\! \bigl(\spc e^{b_+\phi_+} + e^{b_-\phi_-}\spc \bigr)
\eea
This identity is formally implemented by introducing  a lagrange multiplier $\mu_B$  and inserting 
\bea
\int \! d\mu_B \spc e^{2\mu_B(\tau_{2}-\tau_{1})} \exp\Bigl(-\int_{1}^2\!\! \mu_B \spc e^{b_+ \phi_+} - \int_{1}^2\!\! \mu_Be^{b_-\phi_-}\spc \Bigr)
\eea
into the gravitational path integral. Here we have given both Liouville CFTs the same boundary cosmological constant.  From the 3D/2D perspective, this prescription follows from the fact that the 3D dreibein is equal to the difference between the two $SL(2,\mathbb{C})$ gauge fields $e^a = (\Aplus^a - \Aminus^a)/2$. In the 2D/1D correspondence with the double scaled SYK model, this choice implements the equal energy constraint $H_L - H_R = 0$. 

We are thus led to make the following identification between the SYK energy and boundary cosmological constant
\bea
2\mu_B \is E \, = \,  -\frac{2\cos\lambda s }{\sqrt{\lambda (1-q)}}
\eea
Comparing with \eqref{mubrel}, we find an exact match after making the further identifications
\bea 
\label{dictionary}
q\spc =\spc e^{-\lambda}\spc = \spc e^{2\pi i b^2}, \qquad \lambda  \spc =\spc -2 \pi i b^2, \qquad \mu_L \!\is\! \frac{1}{2\lambda}
\eea
These are the anticipated relations derived from our other comparisons between the two systems. We view this match as further strong support for our proposed holographic dictionary.

\section{SYK Correlators from 2D Liouville-de Sitter Gravity}

\vspace{-1mm}

In this section we compute the two-point function of two physical boundary vertex operators ${\cal O}_{\Delta}(\tau)$  in the Liouville-de Sitter gravity model. We make the Ansatz that ${\cal O}_{\Delta}(\tau)$  can be decomposed into the product \eqref{vphys} of two Liouville exponential with momenta $\mfab_+ = b_+\Delta$ and $\mfab_- = b_-(1-\Delta)$. 

\subsection{Canonical Normalization}

\vspace{-1mm}

Before we start the computation, it will be useful to explain our normalization convention. In ordinary 2D CFT, the two-point function of two local operators is usually normalized to one. For boundary Liouville theory, it is customary to rescale operators so that the two point function between $V_\alpha = e^{\alpha\phi}$ and its shadow $V_{Q-\alpha} = e^{(Q-\alpha) \phi}$ is normalized to one. Here we will use a different but equally natural normalization convention, which we call the canonical normalization. Indeed, as we will explain, in our normalization convention the two-point function is in fact {\it invariant} under rescalings of the vertex operators and intermediate states, as long as the rescaling of ${\cal O}_\Delta$ or the intermediate state $|s\ra$ only involve factors that depend on the quantum numbers of the vertex operator or state itself. This canonical normalization is defined as follows.

Consider a two point function between two boundary operators ${\cal O}_{\Delta}$ in Liouville-de Sitter gravity
\bea
G_\Delta(s_0,\tau_1) \is \ \begin{tikzpicture}[scale=0.56, rotate=-90, baseline={([yshift=-.1cm]current bounding box.center)}]
\draw[thick] (0,0) circle (1.5);
\draw[thick,\darkblue] (1.5,.05) -- (.35,.05);
\draw[thick,\darkredd] (1.5,-.05) -- (.35,-.05);
\draw[thick,\darkblue] (-1.5,.05) -- (-.35,.05);
\draw[thick,\darkredd] (-1.5,-.05) -- (-.35,-.05);
\draw[blue,fill=purple] (-1.5,.0) circle (0.12);
\draw (0,1.88) node {\footnotesize $\tau_1$};
\draw[blue,fill=purple] (1.5,0) circle (0.12);
\draw (0.05,-1.88) node {\footnotesize $s_0$};
\draw (-0,0) node {\footnotesize \textcolor{\darkblue}{$\Delta$}};
\end{tikzpicture}\, 
\eea
We claim that this two-point function can always be written in the following canonical form
\bea
\label{generalg}
G_\Delta(s_0,\tau_1) \is \rho(s_\Delta) \rho(s_0) \int\! ds_1 \, \rho(s_1) \, {C}_{\ddelta}(s_0,s_1) \spc {C}_{1-\ddelta}(s_0,s_1)\, e^{-\tau_1\mu_B(s_1)}{}^{{}_{\strut}}_{{}^{\strut}}
\eea
Here $\rho(s)$ denotes the spectral density on the space of physical boundary states $|s\ra$ and ${C}_{\ddelta}(s_0,s_1)$ denotes the square of the matrix element of a boundary vertex operator $V_\Delta$ between two FZZT boundary states labeled by $s_0$ and $s_1$. The above form of the two-point function reflects the fact that physical operators are given by the product of two boundary operators  with conformal dimension $\Delta$ and $1-\Delta$. In \eqref{generalg} it is implied that ${C}_{\ddelta}(s_0,s_1)$ is an analytic function of $\Delta,s_0$ and~$s_1$. 

Neither the spectral density $\rho(s)$ nor the matrix elements ${C}_{\ddelta}(s_0,s_1)$ are uniquely normalized. Changing the normalization convention of states and operators amounts to rescaling the spectral density and matrix elements via 
\bea
\label{rescale}
 \rho(s)\, \to\, \sigma(s) \spc \rho(s),\qquad & & \quad {C}_{\ddelta}(s_0,s_1) \; \to \; \frac{{C}_{\ddelta}(s_0,s_1)}{\sqrt{\sigma(s_\ddelta)\sigma(s_0) \sigma(s_1)}}
\eea
This combined rescaling leaves the two-point function \eqref{generalg} unchanged. Note that the matrix elements ${C}_{\ddelta}(s_0,s_1)$ contain irreducible dependence on the three quantum numbers $\Delta$, $s_0$ and $s_1$ that can not be removed or changed via rescalings of the form \eqref{rescale}. 

The discussion thus far does not yet uniquely fix the two-point function, since we haven't yet indicated how to relate the choice of spectral density $\rho(s)$ to a given normalization of the three point function ${C}_{\ddelta}(s_0,s_1)$. We will now eliminate this ambiguity by imposing the requirement that the spectral density and three point function are directly related  via \cite{Collier:2023fwi}
\bea
\label{limc}
\lim_{\Delta \to 0 } \, {C}_{\ddelta}(s_0,s_1)\! \is \frac{1}{\rho(s_0)}\, \delta(s_1-s_0)
\eea
With this relation, we have uniquely fixed the normalization of the two-point function \eqref{generalg} in terms of the irreducible component of the three point function ${C}_{\ddelta}(s_0,s_1)$. Note that the identity \eqref{limc} is preserved under simultaneous rescalings of the form \eqref{rescale}. To reduce the number of independent quantities to keep track of, we will often use the freedom \eqref{rescale} to impose the special condition that the spectral density $\rho(s)ds = d\mu_B(s)$.

\subsection{Boundary Two-Point Function I}

\vspace{-1mm}

The boundary two point function in Liouville-de Sitter gravity is given by the inner product of the two space-like Liouville boundary two-point functions 
\bea
\label{gzzzz}
G_\Delta(\tau_1\spc) \! \is\!    \int_{\cal M} \!  Z_{\rm gh}\, Z^{+}_\Delta(\tau_1)  \spc {Z^-_{1-\Delta}(\tau_1)}\,=\,
\bigl\langle  Z^{+}_\Delta(\tau_1) \spc \bigl|\spc {Z^-_{1-\Delta}(\tau_1)}\spc \bigr\rangle 
\eea
Here $ \int_{\cal M}$ is a schematic notation for coupling to gravity and performing a proper BRST gauge fixing procedure. The boundary two-point function on the disk doesn't require any moduli integration, but does require gauge fixing one global conformal Killing symmetry. The individual $\phi_+$ and $\phi_-$ boundary two-point functions admit the following decomposition into a sum of conformal blocks
\bea
Z^+_\Delta(\tau_1\spc) \! \is\!  
\int\! d\mu_{\mbox{\tiny$B$}}(s_1)\smpc \spc e^{- \tau_1\mu_B(s_1)}
\ {\rmC^+_{\Delta}(s_0,s_1)} \,  \Psi^{+}_\Delta({s}_0,s_1)
\nonumber \\[-1.5mm] \label{zdeco} \\[-1.5mm]\nonumber
Z^-_{1\nspc-\nspc\Delta}(\tau_1\spc) \!\is\! 
\int\! d\mu_{\mbox{\tiny$B$}}(s_1)\smpc \spc e^{- \tau_1\mu_B(s_1)} \, {\rmC^-_{1-\Delta}(s_0,s_1)} \, {\Psi^-_{1-\Delta}(s_0,s_1)}
\eea
Here ${\rmC^+_{\Delta}}$ and ${\rmC^-_{1-\Delta}}$ denote the square of the matrix elements of the boundary condition changing operators and $\Psi^{+}_\Delta$ and $ {\Psi^-_{1-\Delta}}$ denote the corresponding conformal blocks. 

As outlined in section 3, the conformal blocks of the Virasoro CFTs can be identified with Hilbert states of the 3D de Sitter gravity theory.
As eigen functions of hermitian operators that measure the spectral parameter $s$ of the FZZT state, the blocks are guaranteed to form an orthonormal basis.  We will choose to normalize them as follows.
\bea
\label{specmeasure}
\bigl\langle  \Psi^{+}_\Delta({s}_0,s_1) \spc \bigl|\spc {\Psi^-_{1-\Delta}(s_0,s'_1)} \bigr\rangle\is 
 \delta\bigl(\mu_{\mbox{\tiny$B$}}({s}_1)\! -\! \mu_{\mbox{\tiny$B$}}({s}'_1)\bigr)_{{}^{\strut}}^{{}_{\strut}}
\eea
Plugging in the expansions \eqref{zdeco} gives the boundary two point function \eqref{gzzzz} the anticipated form
\bea
G_\Delta(\tau_1\spc) \! \is\!    \int\! d\mu_{\mbox{\tiny$B$}}(s_1)\smpc \spc e^{- 2\tau_1\mu_B(s_1)}\,{\rmC^+_{\Delta}(s_0,s_1)}\, \rmC^-_{\smpc1\nspc\smpc-\nspc\smpc\Delta}(s_0,s_1)
\eea

We claim that with the above normalization of the blocks, the OPE coefficients ${\rmC^+_{\Delta}(s_0,s_1)}$ and ${\rmC^-_{1-\Delta}(s_0,s_1)}$ coincide with the non-chiral DOZZ three point functions given in equation \eqref{dozzs}. The argument goes as follows. 
Consider the $\tau_1 \to 0$ limit of the boundary two point function
\bea
\label{limz}
\lim_{\tau_1\to 0} Z^+_\Delta(\tau_1) \! \is\!   
\int\!d\mu_{\mbox{\tiny$B$}}(s_1)\smpc\spc 
{\rmC^+_{\Delta}(s_0,s_1)} \,  \Psi^{+}_\Delta({s}_0,s_1)
\eea
The time variable $\tau_1$ represents the proper time distance between the two boundary operators. Hence in the above limit, the two boundary operators are directed to become coincident and act at the same point. It is appropriate to replace the two coincident operators by their dominant OPE channel, given by the identity operator. We deduce that the two point function reduces to the identity conformal block of the form
\bea
\label{id}
\lim_{\tau_1\to 0} Z^+_\Delta(\tau_1) \! \is\!   
\begin{tikzpicture}[scale=0.75, rotate = -90, baseline={([yshift=-0.1cm]current bounding box.center)}]
\tikzset{
    partial ellipse/.style args={#1:#2:#3}{
        insert path={+ (#1:#3) arc (#1:#2:#3)}
    }
} 
\draw[thick] (0,.25) circle (1.1);
\draw[thick,\darkblue] (0,1.35) -- (0,2.25);
 \draw[thick,\darkblue] (0,3) [partial ellipse=-247:63:.75cm and .75cm];
\draw[fill,\darkblue] (0,2.25) circle (0.08);
\draw (0.0,-1.15) node {\small {$s_0$}};
\draw (0.35,1.8) node {\footnotesize  {$\mathbb{1}$}};
\draw[fill,\darkblue] (0,1.35) circle (0.08);
\draw (0.01,3.8) node {\small {\textcolor{\darkblue}{$\Delta$}}};
\end{tikzpicture}
\eea
Here the right-hand side indicates the two-point boundary conformal block projected on the identity channel, as indicated.
Since the conformal blocks $\Psi^{+}_\Delta({s}_0,s_1)$ form a complete basis, we can decompose this block as
\bea
\label{fusion}
\begin{tikzpicture}[scale=0.75, rotate = -90, baseline={([yshift=-0.1cm]current bounding box.center)}]
\tikzset{
    partial ellipse/.style args={#1:#2:#3}{
        insert path={+ (#1:#3) arc (#1:#2:#3)}
    }
} 
\draw[thick] (0,.25) circle (1.1);
\draw[thick,\darkblue] (0,1.35) -- (0,2.25);
 \draw[thick,\darkblue] (0,3) [partial ellipse=-247:63:.75cm and .75cm];
\draw[fill,\darkblue] (0,2.25) circle (0.08);
\draw (0.0,-1.15) node {\small {$s_0$}};
\draw (0.35,1.8) node {\footnotesize  {$\mathbb{1}$}};
\draw[fill,\darkblue] (0,1.35) circle (0.08);
\draw (0.01,3.8) node {\small {\textcolor{\darkblue}{$\Delta$}}};
\end{tikzpicture} \ \ \; \raisebox{0pt}{{\large $= \ \ $}}{\int d\mu_{\mbox{\tiny$B$}}(s_1)\smpc\,  
F^+_{\mathbb{1} s_1} \Bigl[\!\mbox{{$\begin{array}{cc}\!{s}_0\!\!&\!\!\mbox{\scriptsize $\Delta$}\!\\[-1.5mm]\!{s}_0\!\!&\!\!\mbox{\scriptsize $\Delta$}\!\end{array}$}}\!\Bigr]}~~
\begin{tikzpicture}[scale=0.67,rotate=-90, baseline={([yshift=-0.05cm]current bounding box.center)}]
\draw[thick] (0,0) circle (1.25);
\draw[thick,\darkblue] (-1.25,0) -- (-.35,0);
\draw[thick,\darkblue] (.35,0) -- (1.25,0);
\draw[fill,\darkblue] (-1.25,0) circle (0.08);
\draw (0.0,1.6) node {\small {$s_1$}};
\draw[fill,\darkblue] (1.25,0) circle (0.08);
\draw (0.0,-1.6) node {\small {$s_0$}};
\draw (0,0) node {\small \textcolor{\darkblue}{$\Delta$}};
\end{tikzpicture}
\eea
where $F^+_{\mathbb{1} s_1} \Bigl[\!\mbox{{$\begin{array}{cc}\!{s}_0\!\!&\!\!\mbox{\scriptsize $\Delta$}\!\\[-1.5mm]\!{s}_0\!\!&\!\!\mbox{\scriptsize $\Delta$}\!\end{array}$}}\!\Bigr]$ denotes the fusion matrix that relates the identity conformal block to the dual conformal block with intermediate channel $s_1$. 

The above reasoning has become a standard tool in the study of holographic conformal field theory. When applied to the four-point functions on the sphere, the same logic leads to the well-known identification between the above fusion matrices and the DOZZ three point functions
\bea
F^+_{\mathbb{1} s_1} \Bigl[\!\!\spc\mbox{{$\begin{array}{cc}\!{s}_0\!\!&\!\!\mbox{\scriptsize $\Delta$}\!\\[-1.5mm]\!{s}_0\!\!&\!\!\mbox{\scriptsize $\Delta$}\!\end{array}$}}\!\!\Bigr] \!\is \; \rmC^+_{\smpc \ddelta}\smpc(s_{0}, s_1) 
\qquad \qquad F^-_{\mathbb{1} s_1} \Bigl[\!\!\spc\mbox{{$\begin{array}{cc}\!{s}_0\!\!&\!\!\mbox{\scriptsize $1\!\nspc-\!\nspc\Delta$}\! \\[-1.5mm]\!{s}_0\!\!&\!\!\mbox{\scriptsize $1\!\nspc-\!\nspc\Delta$}\!\end{array}$}}\!\!\Bigr]  \,=\,  \rmC^-_{1\nspc-\nspc \Delta}\nspc(s_{0}, s_1),{}_{{}^{\strut}}^{{}_{\strut}}
\eea
where both sides are normalized with respect to the same spectral measure \eqref{specmeasure}.
Hence, by comparing the conformal block decompositions \eqref{limz} and \eqref{fusion} and using the equality \eqref{id}, we can now confirm that the coefficients ${\rmC^+_{\Delta}(s_0,s_1)}$ and ${\rmC^-_{1-\Delta}(s_0,s_1)}$ are given by the DOZZ three point functions \eqref{dozzs} of two space-like Liouville CFTs with total cenrtal charge $26$. 

Using Zamolodchikov's formula \eqref{magicformula} we thus arrive at our final result for the boundary two-point function of Liouville-de Sitter gravity. The S-transformed version of  \eqref{magicformula} reads
\bea
\label{magicformulatwo}
 \CC^+_{\smpc \Delta}\smpc(s_{0}, s_{1})\,  \CC^-_{\smpc1\nspc\smpc-\nspc\smpc\Delta}\nspc(s_{0}, s_{1}\nspc)
\! \is   \frac{\vartheta_1  \bigl( 2b^2 \Delta \qbigrr\spc \prod_a {\vartheta_1  \bigl( 2i b^2 s_{a} \qbigrr} 
 \strut}{\smpc \vartheta_1  \bigl(i b^2 (i\Delta \pm  s_0\pm s_1)\qbigrr}_{{}{\strut}}^{{\strut}}
\eea 
Identifying $\lambda = -2\pi i b^2$ and $q= e^{2\pi i b^2} = e^{-\lambda}$, this matches the two-point function \eqref{gexact} of physical operators in the doubled SYK model. This match is the main result of this paper.

\subsection{Boundary Two-Point Function II}
\vspace{-1mm}

It may be instructive to rederive the result \eqref{gexact} directly from the standard expressions for the boundary Liouville two point function. Conventionally, boundary condition changing operators in Liouville CFT are normalized such that the mixed two-point function between one boundary operator and its shadow is set equal to unity
\bea
\bigl\la B^{s_0s_1}_{\ddelta} B^{s_1s_0}_{\ddelta'} \bigr\ra \! \is \! 
 \; \delta(s_\ddelta + s_{\ddelta'})+ {\cal R}(s_\ddelta|s_0,s_1) \delta(s_\ddelta\!-\nspc s_{\ddelta'})_{{}^{\scriptsize\strut}}^{{}_{\scriptsize \strut}}
\eea
The presence of the second term is a consequence of the reflection property of the associated states. The reflection coefficients in the two space-like Liouville theories take the following form
\bea
{\cal R}_\pm(s_\ddelta|s_0,s_1)  \spc = \spc \frac{
{\rm g}_\pm(-s_{\Delta}|s_{0},s_{1})}
{{\rm g}_\pm(s_{\Delta}|s_{0},s_{1})}
\eea
where\\[-11mm]
\bea
{\rm g}_+(s_\ddelta|s_{0},s_{1})\! \is \,
\frac{\Gamma_{\nspc b}\bigl(\mbox{\small $Q/2$} + is_\ddelta  \pm is_{0}  \pm is_1\bigr)}{\Gamma_{\nspc b}(2is_\ddelta) 
\, \nu(s_0) \, \nu(s_1) 
 \,}
\\[2mm]
{\rm g}_-(s_\ddelta|s_{0},s_{1})\! \is \! 
\frac{\Gamma_{\nspc\nspc ib}\bigl(ib-i\mbox{\small $Q/2$} +s_\ddelta \pm s_{0}  \pm s_1\bigr)}{  \Gamma_{\nspc\nspc ib}(\nspc 2s_\ddelta) 
\, \nu(s_0) \, \nu(s_1) 
\,}
\eea
Here $\Gamma_b(x)$ denotes the Barnes double Gamma function (see Appendix A) and $\nu(s)$ are for now undetermined normalization factors.

Another natural choice is to normalize the boundary operators such that the two-point function takes the symmetric form
\bea
\label{symvv}
\bigl\la V^{s_0s_1}_{\ddelta}V^{s_1s_0}_{\ddelta'} \bigr\ra \! \is \! 
 \rmC_{\Delta}(s_0,s_1)  \Bigl[\delta(s_\ddelta + s_{\ddelta'})+  \delta(s_\ddelta\!-\nspc s_{\ddelta'})\Bigr]_{{}^{\scriptsize \strut}}
\eea
In this normalization, the reflection coefficient that relates the boundary operators to their shadow operators is set equal to 1.\footnote{An analogous normalization was employed in \cite{Fredenhagen:2004cj} to establish a precise match between boundary correlation functions in $c=1$ string theory and the dual realization in terms of matrix quantum mechanics.}
The operators $V^{s_0s_1}_{\ddelta}$ and $B^{s_0s_1}_{\ddelta}$ are related via a rescaling 
\bea
V^{s_0s_1}_{\ddelta} \!\! \is \! g(s_\ddelta|s_0,s_1)\spc B^{s_0s_1}_{\ddelta} 
\eea
with $g(s_\ddelta|s_0,s_1)$ as given above. This redefinition is unique up to the freedom to adjust the normalizations factors $\nu(s_a)$. Modulo this ambiguity, we have thus determined the boundary two point functions in the symmetric normalization \eqref{symvv} \nopagebreak
\bea
\label{twopt}
\rmC^+_{\smpc \Delta}\smpc(s_{0}, s_1) \! \is \! \spc {\rm g}_+(s_{\Delta}|s_{1},s_{0})\spc
{{\rm g}_+(-s_{\Delta}|s_{0},s_{1})} \nonumber\\[-1.5mm]\\[-1.5mm]\nonumber
\rmC^-_{1\nspc-\nspc\Delta}\nspc(s_0,s_1) \! \is \! {\rm g}_-(s_{\Delta}|s_{1},s_{0})\spc{{\rm g}_-(-s_{\Delta}|s_{1},s_{0})}
\eea
For suitable choice of the function $\nu(s)$ these coincide with the DOZZ three point functions.

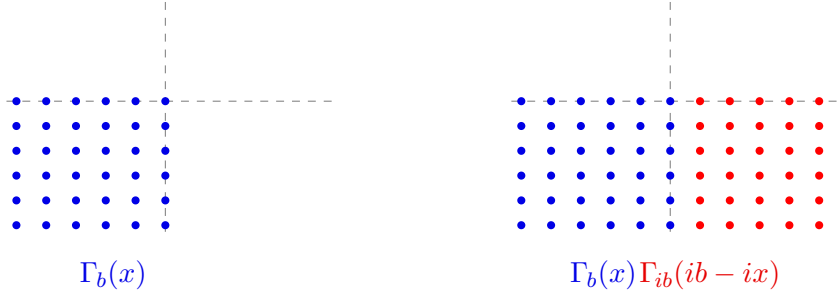
\begin{figure}[t]
\centering
\begin{tikzpicture}[scale=.33,rotate=180]
 \foreach \x in {6} {
\draw[gray, dashed, thin] (1.2 * \x,2) -- (1.2 * \x,11.5);
}
 \foreach \y in {6} {
 \draw[gray, dashed, thin] (.5,\y) -- (13.8,\y);
}  
\foreach \x in {6,7,8,...,11} {
    \foreach \y in {6,7,8,...,11} {
      \draw[thick,fill=\darkblue,\darkblue] (1.2 * \x,\y) circle (0.12); 
    }
    } 
    \draw (9.3,13) node {\textcolor{\darkblue}{$\Gamma_{\nspc b}(x)$}};
    \end{tikzpicture}~~~~~~~~~~~~~~~~~~\begin{tikzpicture}[scale=.33,rotate=180]
   \foreach \x in {6} {
 \draw[gray, dashed, thin] (1.2 * \x,2) -- (1.2 * \x,11.5);
}
 \foreach \y in {6} {
 \draw[gray, dashed, thin] (.5,\y) -- (13.8,\y);
} 
    \foreach \x in {6,7,8,...,11} {
    \foreach \y in {6,7,8,...,11} {
      \draw[thick,fill=\darkblue,\darkblue] (1.2 * \x,\y) circle (0.12); 
    }
    }
\foreach \x in {1,2,...,5} {
    \foreach \y in {6,7,...,11} {
      \draw[thick, fill=\darkredd,red] (1.2 * \x,\y) circle (0.12); 
    }
    }
        \draw (7,13) node {\textcolor{\darkblue}{$\Gamma_{\nspc b}(x)\spc$}\textcolor{\darkredd}{$\Gamma_{\nspc\nspc ib}(ib-ix)$}};
\end{tikzpicture}
\vspace{-1.5mm}
\caption{The product of two $\Gamma_b$ functions has the same set of poles as the q-Gamma function~$\Gamma_{\tilde{q}}$.}
\vspace{-2mm}
\end{figure}

The derivation thus far does not prescribe the spectral measure by which to integrate over the intermediate FZZT variable $s_1$.  However, we can determine this spectral measure by applying our earlier discussion of the canonical normalization of the two-point function $G_\Delta(s_1)$. In addtion, we can use the following identity between the product of two Barnes double Gamma functions and the q-deformed Gamma function
\bea
\label{gbq}
\Gamma_b(x) \Gamma_{ib}(ib-ix) \!\is \! \frac{(-\sqrt{\Tilde{q}},\Tilde{q})_\infty}{(\Tilde{q},\Tilde{q})_\infty}\spc \frac{e^{-\frac{i\pi}{4}(\mbox{\small $Q/2$} -x)^2}} {(1-\tilde{q})^{1-x}}\; {\Gamma_{\tilde{q}}(b\smpc x)}
\eea
with $\tilde{q} = e^{2\pi i/b^2}$. This identity fits the same pattern as the previous ones: the function $\Gamma_b(z)$ has poles for $z=-nb-{m}{b}^{-1} $ with $m,n\in~\mathbb{Z}^+$ with 
$\Gamma_{\tilde{q}}(z)$ has poles for $z=-nb-{m}{b}^{-1} $ with $m\in~\mathbb{Z}^+$ and $n \in \mathbb{Z}$. Figure 6 shows a pictorial representation of this identity.

Applying \eqref{gbq} to the product of the boundary two point functions \eqref{twopt}, and making a judicial choice of the $\nu(s)$ factors, yields the following identity
\bea
{\rm g}_+(s_{\ddelta}|s_{0},s_{1})\spc
{{\rm g}_+(-s_{\ddelta}|s_{1},s_{0})} \spc  {\rm g}_-(s_{\ddelta}|s_{0},s_{1})\spc{{\rm g}_-(-s_{\ddelta}|s_{1},s_{0})}  \! \is \! {\bf C}_{\smpc \Delta}\smpc(s_0,s_1)\spc {\bf C}_{1\nspc-\nspc\Delta}\nspc(s_0,s_1) \quad
\eea
where\footnote{Here we made advance use of the $b \to 1/b$  invariance of the total expression to replace $\Gamma_{\tilde{q}}$ by $\Gamma_q$.}  \\[-9mm]
\bea
\label{cspecial}
{ \bf C}_{\smpc \Delta}\smpc(s_{0}, s_1)\is 
\, \frac{\Gamma_{\nspc {q}}\bigl(\Delta  \pm i{s}_{0}  \pm i{s}_1\bigr)}{ 
|\Gamma_{\nspc {q}}(2\Delta){\Gamma_q(2is_0) \, \Gamma_q(2i s_1)}|}\, .
\eea
The expression in the numerator is irreducible and independent of normalization convention. The expression in the denominator is adjusted such that
\bea
\lim_{\Delta \to 0 } \, {\bf C}_{\ddelta}(s_0,s_1) \! \is \delta(s_1-s_0)\, .
\eea
Hence, according to our canonical normalization prescription, the spectral density is set to a constant. We thus obtain the following result for the boundary two-point function 
\bea
G_\Delta(\tau_1)  \is \, \int\! ds_1 \, {\bf C}_{\ddelta}(s_0,s_1) \spc {\bf C}_{1-\ddelta}(s_0,s_1)\, e^{-\tau_1\mu_B(s_1)} \,.
\eea
Upon inserting \eqref{cspecial}, this again reproduces the SYK result \eqref{gexacto}.



\section{Conclusion}

\vspace{-1.5mm}

\noindent
We have investigated the relationship between three exactly solvable low dimensional quantum systems, the 1D double scaled SYK model, 3D de Sitter gravity, and a new 2D gravity theory consisting of two spacelike Liouville CFTs with complex central charge $c_\pm = 13 \pm i\gamma$ and total central charge $c_++c_-=26$. Motivated by this triangle of relationships, we studied the boundary two-point function of physical operators in the Liouville gravity system as a function of the geodesic distance and found a precise match with the two-point function of physical operators in the doubled DSSYK model studied in \cite{ustwo}. As we explained, to obtain this match, it was necessary to  consider a two-sided DSSYK system and combine both space-like Liouville CFTs into a single gravity theory. This correspondence indicates that the spacelike-Liouville gravity model is well-defined and unitary. 

One obvious question is whether a 2D gravity model based on CFTs with a complex central charge can really be a unitary quantum system, since the individual Virasoro modules necessarily contain negative norm states \cite{Balasubramanian:2002zh}. One natural definition of hermitian conjugation is $(L_n^-)^\dag = L_{-n}^+$  and $(\tilde{L}_n^-)^\dag =  \tilde{L}_{-n}^+$. With this choice, pure $L_{-n}^+$ or $L_{-n}^-$ descendants have zero norm and general linear combinations have either positive or negative norm. To obtain the physical theory, we must add a ghosts and impose BRST invariance and the general reasoning of \cite{Spiegelglas:1986xe} demonstrates that for worldsheet CFTs with total central charge $26$, all such negative norm states are removed after imposing the physical state conditions. The resulting physical spectrum then consists of only primary states and all have positive norm.

Another question that requires more study is whether the Liouville-de Sitter gravity allows for other boundary states besides the FZZT states whose boundary two-point function we investigated in section 5. Any consistent boundary state would have to be a linear combination of boundary states $|\!|\alpha\ra\!\ra$ defined via the usual Ishibashi conditions
$(L_n^+ - \tilde{L}^+_{-n}) |\!|{\rm \alpha}\ra\!\ra = (L_n^- - \tilde{L}_{-n}^-) |\!|{\rm \alpha}\ra\!\ra = \, 0.$ Perhaps the most surprising conclusion of our study is that the boundary cosmological constant $\mu_B$ that specify the FZZT states takes values over a finite range. Correspondingly, since FZZT states are labeled by means of the state running in the dual channel, we are led to conclude that the spectrum of primary states of physical states in the spacelike-Liouville gravity model also runs over a finite range. This type of spectrum is radically different from that of conventional CFTs, which typically contains states with arbitrarily high energy. The bounded spectrum of the 2D gravity model is the strongest evidence that it should be viewed as a description of de Sitter space, as opposed to anti-de Sitter space. 

In this paper, we outlined how the 2D spacelike-Liouville gravity model arises from quantization of 3D de Sitter gravity. The derivation made use of a `quantize first and impose physical constrains afterwards' approach and naturally leads to a description in terms of 2D LCFT subject to physical state conditions that implement 2D diffeomorphism invariance. In the companion paper \cite{Verlinde:2024znh}, the opposite `constrain first and then quantize' method is studied and found to lead to a direct link between 3D de Sitter gravity and the combinatorial rules that govern the exact computation of the spectrum and correlation function in double scaled SYK. The approach taken in \cite{Verlinde:2024znh} is to directly quantize the moduli space of non-rotating Schwarzschild-de Sitter spacetimes. The Hamiltonian is identified with the gravitational Wilson-line that measures the conical deficit angle. It should be possible to give a parallel derivation of this correspondence by studying the action of the Verlinde loop operators of (2,1) degenerate fields \cite{Gaiotto:2014lma} on the space of FZZT boundary states.

\def\mfa{\mbox{\large $\mathfrak{a}$}}
\def\mfaa{\alpha} 
\def\mfb{\mbox{\large $\mathfrak{b}$}}
\smallskip

\smallskip

\section*{Acknowledgments}

We thank Akash Goel and Vladimir Narovlansky for initial collaboration and Andreas Blommaert, Scott Collier, Lorenz Eberhardt, Ping Gao, Beatrix M\"uhlmann, Alex Maloney, Thomas Mertens, Douglas Stanford, and Erik Verlinde for helpful discussions and comments. The research of HV is supported by NSF grant PHY-2209997.

\appendix

\section{Special functions}
In this section, we introduce the definitions of various special functions used in the main text and discuss some useful properties and identities of these functions.

\medskip

\noindent
\textit{Double gamma function }
We start with the Barnes double gamma function $\Gamma_2(x|a_1,a_2)$ \cite{Barnes}, which is closely related to the double gamma function $\Gamma_b(x)$ that appear in the paper. It is a generalization of Euler gamma function, which is a meromorphic function with poles at $x = -ma_1-na_2$ for non-negative integers $m,n$. The Barnes double gamma function admits the following infinite product representation
\begin{equation}
    \begin{aligned}
        \Gamma_2(x|a_1,a_2) =&\frac{e^{-x\underset{s=1}{\text{Res}_0}\zeta_2(s,0|a_1,a_2)+\frac{1}{2}x^2 (\underset{s=2}{\text{Res}_0}\zeta_2(s,0|a_1,a_2)+\underset{s=2}{\text{Res}_1}\zeta_2(s,0|a_1,a_2))}}{x}\\&\times\prod_{(m,n)\neq (0,0)}\frac{e^{\frac{x}{ma_1+na_2}-\frac{x^2}{2(ma_1+na_2)^2}}}{1+\frac{x}{ma_1+na_2}}, 
    \end{aligned}
\end{equation}
where $\text{Res}_n$ takes the n-th order residue and the Barnes zeta function $\zeta_2(s,x|a_1,a_2)$ is defined by the following infinite sum
\begin{equation}
    \zeta_2(s,x|a_1,a_2) = \sum_{m,n\ge 0}\frac{1}{(x+ma_1+na_2)^s}.
\end{equation}
The pole structures of Barnes double gamma function $\Gamma_2(x|a_1,a_2)$ can be read directly from the infinite product representation.

The Barnes double gamma function $\Gamma_b(x)$ mentioned in the main text can be viewed as the normalized $\Gamma_2(x|b,b^{-1})$
\begin{equation}
    \Gamma_b(x) = \frac{\Gamma_2(x|b,b^{-1})}{\Gamma_2(\frac{Q}{2}|b,b^{-1})}. 
\end{equation}
where we define $Q = b+b^{-1}$ and by normalization we have $\Gamma_b(\frac{Q}{2})=1$. Besides the infinite product formula, the double gamma function $\Gamma_b(x)$ also has the following integral representation:
\begin{equation}
    \log \Gamma_b(x) = \int_0^\infty \frac{dt}{t}\left(\frac{e^{-xt}-e^{-Qt/2}}{(1-e^{-bt})(1-e^{-t/b})}-\frac{(Q-2x)^2}{8 e^t}-\frac{Q-2x}{2t}\right)
\end{equation}
Similar to $\Gamma_2(x|a_1,a_2)$, $\Gamma_b(x)$ is a meromorphic function in $x$ and has poles at $x = -mb-nb^{-1}$ for integers $m,\,n \ge 0$. Meanwhile, it has the following shift identity \begin{equation}
    \Gamma_b(x+b) = \sqrt{2\pi}b^{bx-\frac{1}{2}}\Gamma^{-1}(bx)\Gamma_b(x)
\end{equation}
\medskip

\noindent
\textit{$\Upsilon_b$-function}
Given the definition of Barnes double gamma function, the Upsilon function $\Upsilon_b(x)$ shown in the DOZZ formula can be written as the inverse of the product of two double gamma function
\begin{equation}
    \Upsilon_b(x) = \frac{1}{\Gamma_b(x)\Gamma_b(Q-x)}.
\end{equation}
The pole structures of double gamma function implies that the Upsilon function $\Upsilon_b(x)$ is an entire function with zeroes at $x = -mb-nb^{-1}$ and $x = (m'+1)b+(n'+1)b^{-1}$ for integers $m,\,n,\,m',\,n'\ge 0$. 

The Upsilon function can also be defined by the integral formula
\begin{equation}
    \log \Upsilon_b(x) =  \int_0^\infty \frac{dt}{t}\left[ \left(\frac{Q}{2}-x\right)^2e^{-t}-\frac{\sinh^2{\frac{1}{2}\left(\frac{Q}{2}-x\right)t}}{\sinh{\left(\frac{1}{2}bt\right)}\sinh{\left(\frac{1}{2}b^{-1}t\right)}}\right].
\end{equation}
Since it is defined in terms of double gamma function, the Upsilon function possesses the similar shift identity
\begin{equation}
    \Upsilon_b(x+b) = \frac{\Gamma(bx)}{\Gamma(1-bx)}b^{1-2bx}\Upsilon_b(x).
\end{equation}
Note that both $\Gamma_b(x)$ and $\Upsilon_b(x)$ are invariant under $b\leftrightarrow b^{-1}$.

\medskip

\noindent
\textit{$q$-gamma function}
The q-gamma function is a q-analogue of the ordinary gamma function, which mimics the gamma function and has the following shift identity \cite{Askey}
\begin{equation}
    \Gamma_q(x+1) = [x]_q\Gamma_q(x).
\end{equation} $[x]_q$ is the q-number of $x$. In $q\rightarrow 1$ limit, we have $\lim_{q\rightarrow 1}\Gamma_q(x) = \Gamma(x)$. The q-gamma function can be defined by the following infinite product formula 
\begin{equation}
    \Gamma_q(x) = (1-q)^{1-x}\frac{(q;q)_\infty}{(q^x;q)_\infty} = (1-q)^{1-x}\prod_{n=0}^\infty\frac{1-q^{n+1}}{1-q^{n+x}},
\end{equation}
where $(x;q)_\infty = \prod_{n=0}^\infty(1-xq^n)$ is the q-Pochhammer symbol. From the product formula, we find that the q-gamma function has poles at $x = -n-m(2\pi i /\log{q})$ for integer $n\ge 0$ and $m\in\mathbb{Z}$.

The q-gamma function is shown to be closely related to the Barnes double gamma function in \cite{Fredenhagen:2004cj}. By setting $\Tilde{q} = e^{2\pi i/b^2}$, the two functions are connected via the following identity
\begin{equation}
    \Gamma_b(x)\Gamma_{ib}(ib-ix) = e^{-\frac{\pi i}{4}(Q/2-x)^2} \frac{(-\sqrt{\Tilde{q}},\Tilde{q})_\infty}{(\Tilde{q},\Tilde{q})_\infty} (1-\Tilde{q})^{x-1}\Gamma_{\Tilde{q}}(bx).
\end{equation}

\medskip

\noindent
\textit{Jacobi-theta function $\vartheta_1(x,q)$}
The Jacobi-theta function $\vartheta_1(x,q)$ can be defined by the following infinite sum
\begin{equation}
    \vartheta_1(x,q) = i\sum_{n=-\infty}^\infty (-1)^n q^{\frac{1}{2}(n+\frac{1}{2})^2}e^{2\pi i x(n+\frac{1}{2})},
\end{equation}
which has the double quasi-periodicity
\begin{equation}
    \vartheta_1(x+1,q) = -\vartheta_1(x,q),\quad \vartheta_1(x+\tau,q) = -e^{-2\pi i x-i\pi \tau}\vartheta_1(x,q).
\end{equation}
Here $q = e^{2\pi i \tau}$. The Jacobi-theta function has no poles but has zeroes at $x = m+n\tau$ for $m,n \in \mathbb{Z}$. Meanwhile, the Jacobi-theta function has the following modular properties
\begin{equation}
    \vartheta_1(-x/\tau,e^{-2\pi i/\tau}) =e^{\frac{\pi i}{4}}\sqrt{\tau}e^{i\pi x^2/\tau}\vartheta_1(x,e^{2\pi i \tau}). 
\end{equation}

The product of two Upsilon functions $\Upsilon_b(bx)$ and $\Upsilon_{ib}(ib-ibx)$ can be rewritten in terms of the Jacobi-theta function by using the identity proven in \cite{Zamolodchikov:2005fy}
\begin{equation}
    \Upsilon_b(bx)\Upsilon_{ib}(ib-ibx)=e^{i\pi (b-b^{-1}-2bx)^2/8}e^{-i\pi/4b^2}\frac{\vartheta_1(x,\Tilde{q})}{\vartheta_3(0,\Tilde{q})},
\end{equation}
where $\Tilde{q} = e^{2\pi i/b^2}$ and $\vartheta_3(x,q) = \sum_{n=-\infty}^\infty q^{n^2/2}e^{2\pi in x}$.


\def\G{\Gamma}
\def\t{\tau}
\def\tq{\tilde q}
\def\U{\Upsilon}
\def\CP{\mathcal P}
\def\pd{\partial}
\def\Z{{\mathbb{Z}}}
\def\C{{\mathbb{C}}}
\def\CL{{\mathcal L}}
\def\CH{{\mathcal H}}
\def\Li{{\rm Li}}
\def\R{{\mathbb{R}}}

\medskip

\bibliographystyle{ssg}
\bibliography{Biblio}

\end{document}